\documentclass[[journal]{IEEEtran}
\usepackage{cite}
\usepackage{lipsum}
\usepackage{amsmath,amssymb,amsfonts}
\usepackage{algorithmic}
\usepackage{graphicx}
\usepackage{textcomp}
\usepackage{float}
\usepackage{subfloat}
\usepackage{placeins}
\usepackage{booktabs, makecell}
\usepackage{siunitx}
\usepackage{longtable}
\usepackage{hyperref}
\usepackage{caption}
\usepackage{subcaption}
\usepackage{multirow}
\usepackage{multicol}
\usepackage{booktabs}
\usepackage{tabularx}
\newcolumntype{L}{>{\raggedright\arraybackslash}X}
\usepackage{stfloats}   
\usepackage{stfloats}

\begin{document}
\title{2x2 MIMO Prototype for BER and EVM Measurements in Metal Enclosure}
\author{\IEEEauthorblockN{Mir~Lodro\IEEEauthorrefmark{1},~\IEEEmembership{Member,~IEEE,} Gabriele~Gradoni\IEEEauthorrefmark{1},~\IEEEmembership{Member,~IEEE,} Christopher~Smartt\IEEEauthorrefmark{1}, David~Thomas\IEEEauthorrefmark{1}, Steve~Greedy\IEEEauthorrefmark{1}}\\\
\IEEEauthorblockA{\IEEEauthorrefmark{1}{University of Nottingham, NG7 2RD,  United Kingdom}}
\vspace{-0.75cm}
\thanks{Mir Lodro, Steve Greedy, Christopher Smartt, David Thomas and Gabriele Gradoni are with George Green Institute for Electromagnetic Research-GGIEMR, the University of Nottingham, UK. Gabriele Gradoni is also with University of Cambridge, UK.}}
\maketitle
\begin{abstract}
In this work, we present a 2x2 near-field multi-input multiple-output (MIMO) prototype for bit-error-rate (BER) and error vector magnitude (EVM) measurements in a metal enclosure. The near-field MIMO prototype is developed using software-defined-radios (SDRs) for over-the-air transmission of QPSK modulated baseband waveforms. We check the near-field MIMO BER and EVM measurements in three different scenarios in a highly reflecting metal enclosure environment. In the first scenario, the line-of-sight (LOS) communication link is investigated when the mode-stirrer is stationary. In stationary channel conditions near-field MIMO BER and EVM measurements are performed. In the second scenario, BER and EVM measurements are performed in dynamic channel conditions when the mode-stirrer is set to move continuously. In the third scenario, LOS communication near-field MIMO BER and EVM measurements are performed in stationary channel conditions but now in the presence of MIMO interference. In three different scenarios, near-field MIMO BER and EVM measurements are investigated at different Tx USRP gain values and in the presence of varying levels of MIMO interference. 
\end{abstract}

\vspace{-0.1cm}
\begin{IEEEkeywords}
,Software-defined-radio,USRP, EVM, near-field, Rich Scattering,BER measurement, EVM measurement.
\end{IEEEkeywords}

\vspace{-0.35cm}
\section{Introduction}
\label{section label}
Wireless communication system deployment in a highly reflecting and complex propagation environments such as in a metal enclosure or in a compact box environment is extremely challenging \cite{lodro2021near}\cite{lodro2020near}. Near-field MIMO for high-data rate communication is becoming important for applications like wireless chip-to-chip communication and wireless network-on-chip (WNoC) \cite{shamim2016wireless,chen2007inter,lodro2020near,radi2020demonstration,timoneda2018channel, phang2018near}. Traditionally, near-field communication is used for short-range, low-rate and power-constraint application using magnetically coupled loop antennas\cite{kim2017review}. Near-field magnetic induction (NFMI) communication also uses magnetically coupled loop antennas for data transfer\cite{kim2016near}. However, such near-field communication systems operate in high permittivity medium such as underground \cite{guo2020performance}\cite{tan2015wireless}, biological masses \cite{wen2021channel} and underwater environment\cite{akyildiz2015realizing}\cite{li2019survey}. Near-field communication in such environment are sensitive to alignment errors, and offer limited data-rate. This scenario represents a complex propagation environment where in addition to LOS there exists rich-scattering as well as forward propagation through high permittivity materials. Near-field MIMO digital communication system design in such environment is considerably challenging and exposes the benefits of MIMO signal processing \cite{mikki2020theory}. There are channel modeling and characterization efforts for applications like wireless chip-to-chip communication, however, end-to-end near-field MIMO BER performance is not demonstrated. Rx power measurements in the near-field and the path-loss measurements are very rare in the literature. There is significant gap on the assessment of the effect of near-field coupling on the performance of wireless communication system. We check the performance of near-field MIMO communication system in terms of EVM and BER measurements in various scenarios in stirred metal enclosure environment. The near-field MIMO communication scenarios in presence of rich-scattering may arise in different applications such as industry environment, animal cages \cite{sharma2019wideband}, ICT equipment \cite{ohira2011experimental}, Kiosks \cite{he2017stochastic}, computer chassis \cite{redfield2011understanding}, metal cabinets \cite{khademi2015channel} and wireless inter-board communication\cite{karedal2007characterization}\cite{gelabert2011experimental}.
We previously presented OTA testing of QPSK receiver for the performance of baseband receiver design stages and the measurements were performed in a metal enclosure with all baseband receiver design stages \cite{lodro2020near}. In \cite{lodro2021near}, we conducted near-field EVM measurements in a metal enclosure using a single transmit and a single receive antenna. We transmitted non high throughput (nonHT) IEEE 802.11a OFDM packets for all modulation and coding techniques (MCS0-MCS7) in metal enclosure in different loading conditions. The measurement were performed with single channel full-duplex PlutoSDR from Analog Devices. In contrast this work uses two channel and a high performance full-duplex X310 USRP from Ettus Research for EVM and end-to-end BER measurements in metal enclosure in different channel conditions. Based on the application requirements X310 USRP can be interfaced with different RF cards such WBX and UBX-160. We used two UBX-160 per X310 USRP which can offer effective bandwidth of 160 MHz.
 We have organized our paper into five sections. Section I is an introduction that highlights the near-field measurements in metal enclosure. Section II is about X310 USRP based 2x2 near-field MIMO measurement prototype which discusses experimental setup including USRPs, metal enclosure, the type of antennas, and the mode-stirrer. Section III is about the measurement process and the important prototype development steps and measurement stages. Section IV is about experimental results in four different metal enclosure scenarios. It contains a detailed discussion about the near-field MIMO BER and EVM measurements and presents two-channel measurement data. Section V concludes the study and presents future extensions of the prototype and recommendations about near-field measurement scenarios.
\vspace{-0.35cm}
\section{Measurement Setup}
\label{sec:measurement_setup}
\begin{figure}
    \centering
    \includegraphics[width=0.8\columnwidth]{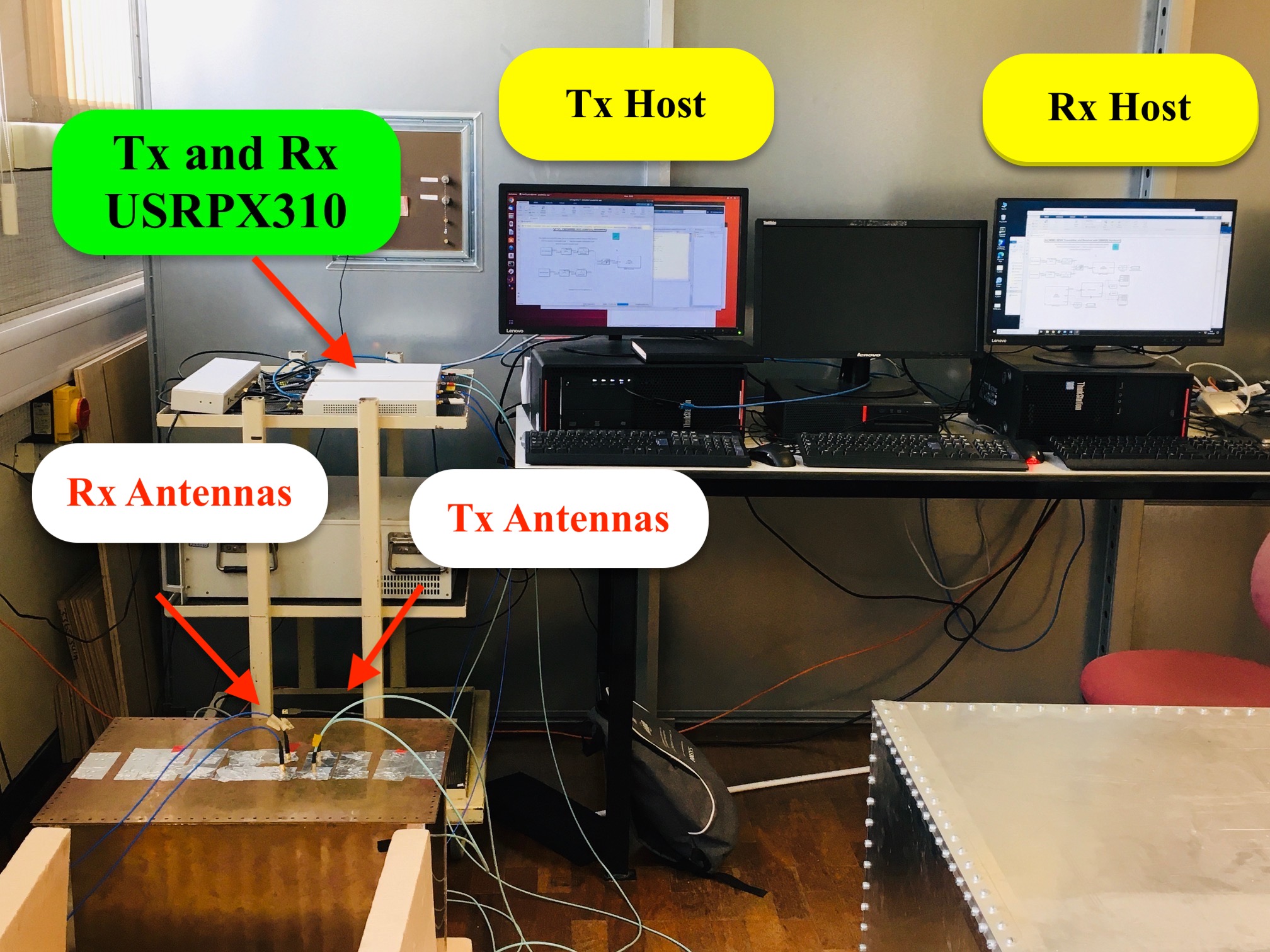}
    \caption{Overall USRP-based 2x2 near-field MIMO measurement setup for BER and EVM measurement.}
    \label{fig:overallmeasurementsetup}
\end{figure}

The measurement setup consists of two X310 USRPs, four wideband antennas, a mode-stirrer, and a high-performance Intel Xeon PC with P400-quadro NVIDIA GPU. We have used a brass metal enclosure with dimensions of $h\times \ell \times w$ of $45\,cm\times 37\,cm\times 55\,cm$. The metal enclosure is a perfectly reflecting environment and when excited at 5.6 GHz it produces rich multipath fading environment.
Near-field MIMO BER and EVM measurement is performed using wideband antennas at frequency of 5.6 GHz in the metal enclosure. The distance between transmitter and receiver $D$ is kept fixed at $50\, mm$ and the distances between transmit antenna elements $d_t$ and receive antenna elements $d_r$ is also kept fixed at $d_t=d_r=45\, mm$ for all RF measurements. The overall MIMO measurement setup for near-field BER and EVM measurement is shown in Fig. \ref{fig:overallmeasurementsetup} where vertically polarized ultra wideband (UWB) transmit and receive antennas are mounted on the metal lid of the enclosure such that Tx and Rx antennas face each other. This scenario depicts a confined propagation environment where Tx and Rx are deployed in the near-field expecting direct coupling as well as rich-scattering introduced by a metal enclosure. Tx and Rx UWB antennas are connected to Tx and Rx X310 USRPs. Data transfer between the host PC and the USRPs takes place over 1 Gigabit Ethernet cable. Real-time streaming of the IQ data is important for EVM and end-to-end BER performance and check the effect of channel in configuration. The presence of any dropped samples i.e. overflows at the receiver introduces inconsistent and intractable performance. In our measurement, we didn't observe any underflow and overflow when the Tx and Rx ports were configured to transfer IQ data to the host PC at sampling rates of 400 ksps. Therefore, the two channel EVM and the BER were function of hardware-impairments and the channel conditions in the enclosure.
Fig. \ref{fig:overallmeasurementsetup} shows the overall measurement setup which consists of two host PCs, two USRPs, Octoclock-G, metal enclosure, and ultra-wideband antennas.

\begin{figure}
    \centering
    \subfloat[]{\includegraphics[width=0.48\columnwidth]{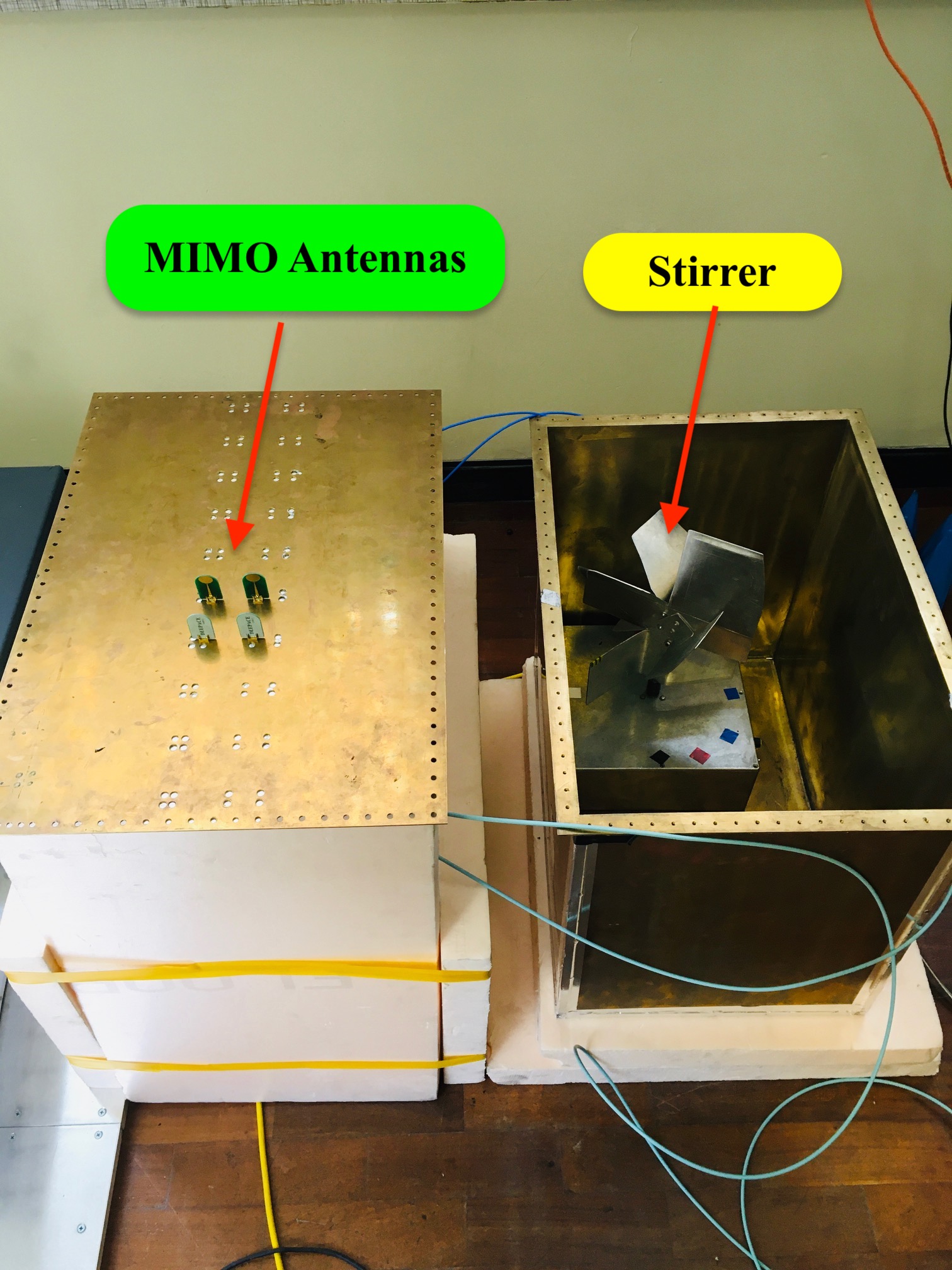}}\hfill
    \subfloat[]{\includegraphics[width=0.48\columnwidth]{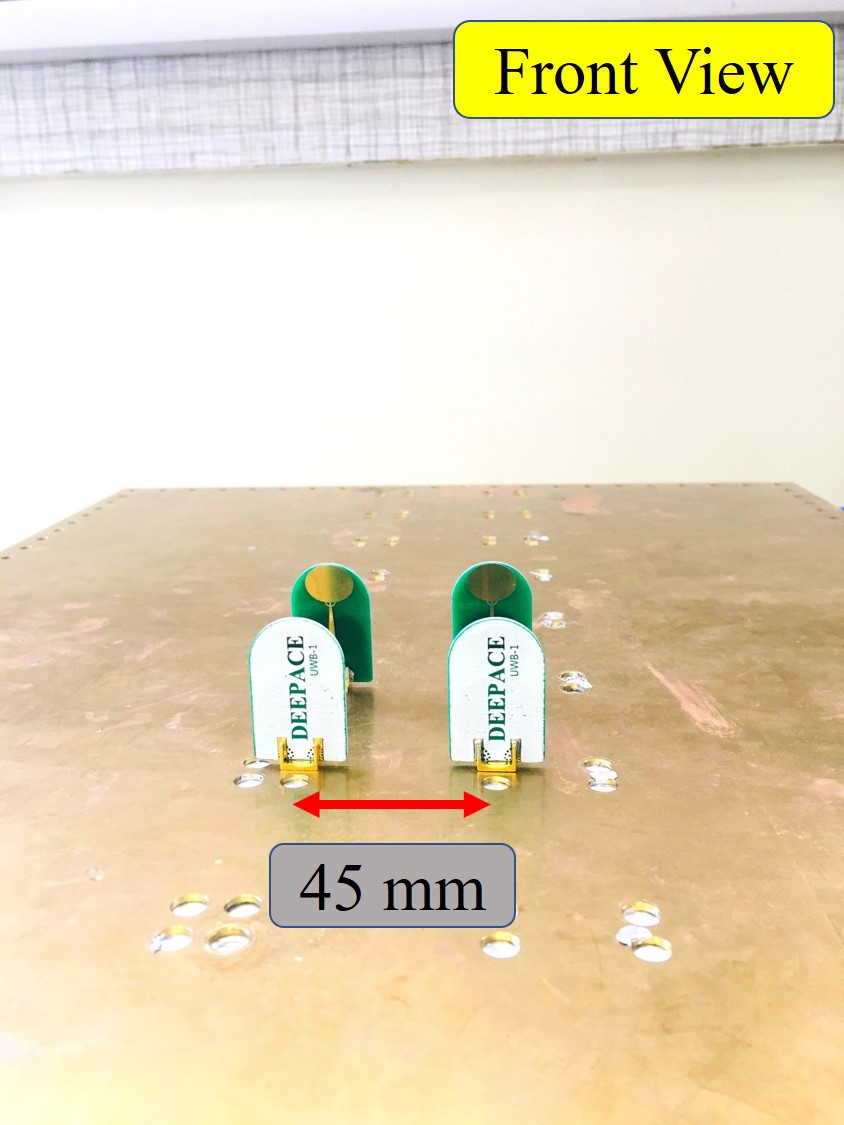}}\\
     \subfloat[]{\includegraphics[width=0.48\columnwidth]{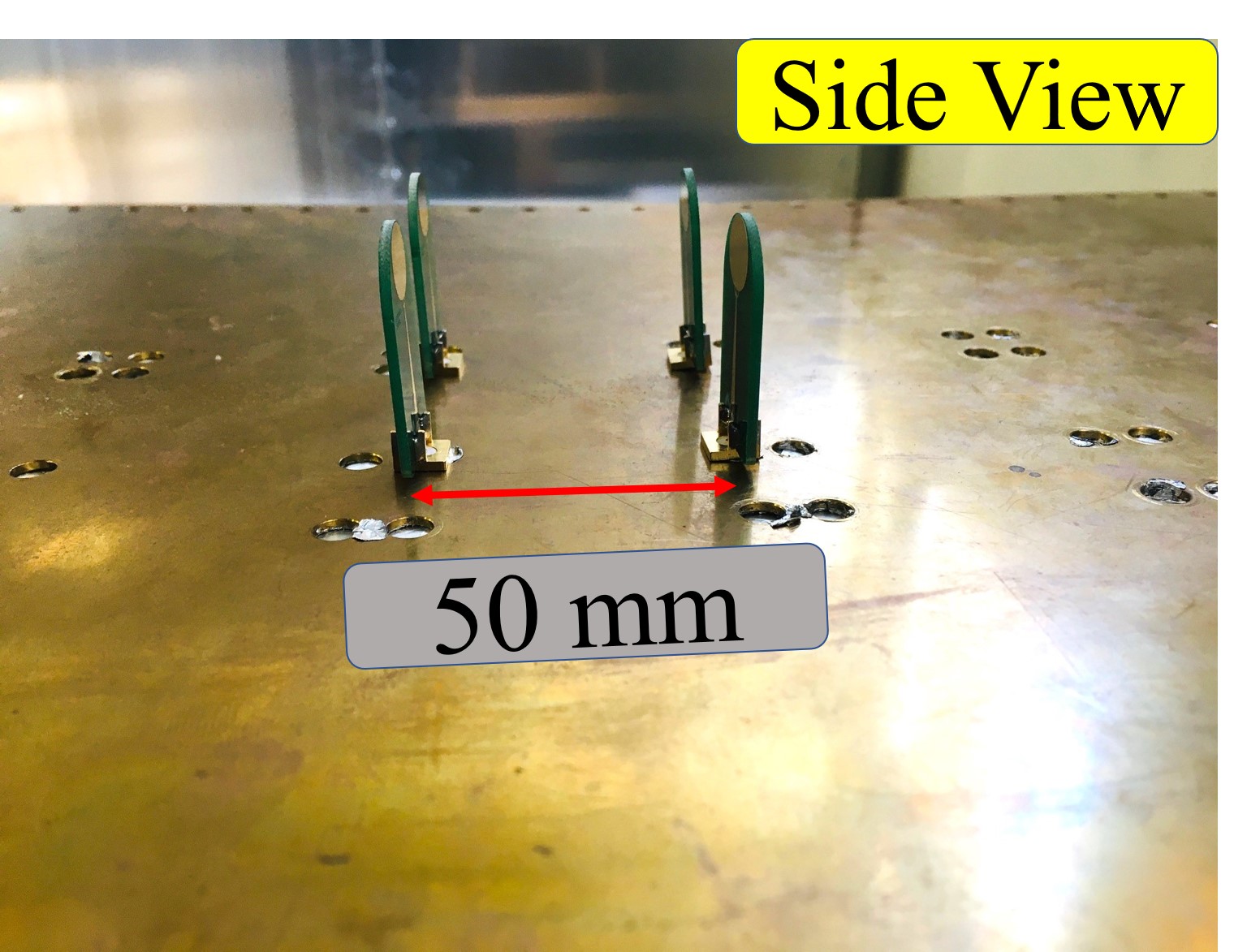}}\hfill
    \subfloat[]{\includegraphics[width=0.48\columnwidth]{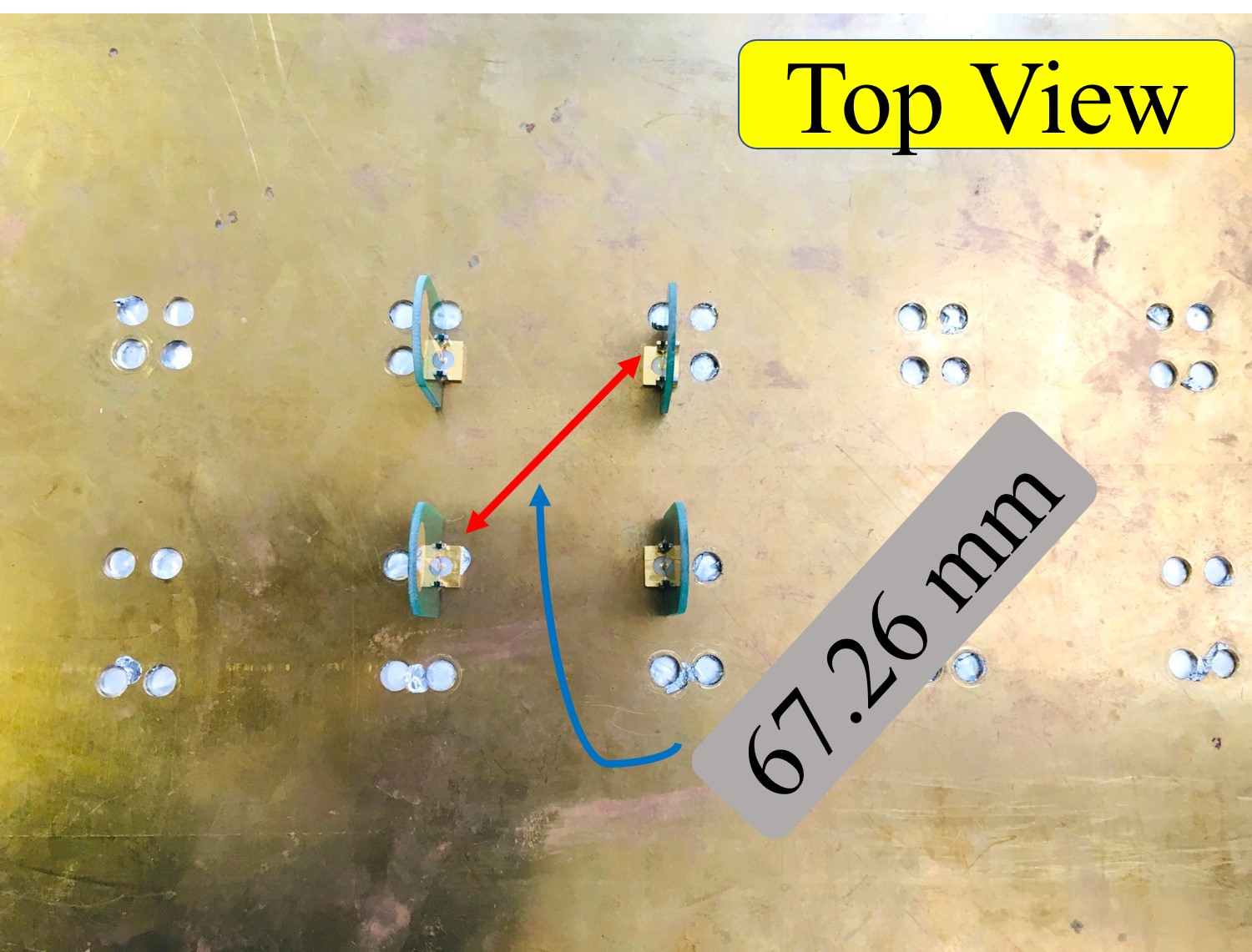}}
    \caption{2x2 MIMO BER and EVM measurement setup inside cavity (a) MIMO antennas and inside view of cavity (b) front view of antennas (c) side view of antennas (d) top view of antennas.}
    \label{fig:insidecavity_views}
\end{figure}
Fig. \ref{fig:insidecavity_views} shows an internal view of the metal enclosure including a stirrer, UWB antennas attached to the metal enclosure lid. This also shows the front-view, side-view, and top-view of broadband MIMO antenna configurations. Holes in the metal panel other than used for different Tx and Rx inter-element spacing created for convenience have been covered with metal tape for RF shielding so that the measurements are not affected by external RF signals.
BER measurement is performed using the QPSK receiver. Before running the model USRPs were frequency calibrated. Frequency offset of 17.19 Hz was found using frequency calibration transmitter and receiver baseband models which is tractable by coarse frequency compensation block in the baseband receiver.
The transmit baseband models consists of frame generation which contains upsampled Barker-13 code as frame header plus payload corresponding to ASCII equivalent of '$\mathrm{Hello\, World}\#\#\#$' where $\#\#\#$ is repeating sequence from 001 to 099. The generated frames are QPSK-modulated with Gray mapping. The QPSK-modulated symbols are upsampled by transmit raise cosine filter producing baseband rate of 400 ksps at the transmitter. Transmit USRP sample rate should match with baseband sampling rate by setting DAC clock rate of 200 MHz with interpolation factor of 500 or DAC clock rate of 20 MHz and interpolation factor of 100. In nutshell, USRP sampling rate is selected as:
\begin{equation}
    \mathrm{Sampling\, Rate=\frac{DAC \left(ADC\right)\, Clock\, Rate}{Interpolation\, \left(Decimation\right)}}
\end{equation}
Over-the-air transmitted QPSK waveform is received by Rx USRP with sample rate of 400 ksps for the baseband processing that takes place in MATLAB session. The baseband QPSK receiver consists of different stages for BER measurements and the information recovery. The baseband receiver stages are AGC, coarse-frequency compensation stage, fine-frequency compensation stage, timing recovery, frame synchronization, phase-ambiguity  resolution, and data-recovery stage. We measured EVM after the fine-frequency stage where the received signal is compensated for negative or positive frequency offsets.  BER measurement is critical KPI for complete end-to-end system performance. The QPSK baseband transmitter and receiver Simulink models are configured for EVM and BER conditions in a MATLAB script. The Tx gain of the USRP changes after the current iteration of BER and EVM measurement is finished. The BER and EVM measurements ares stored in a database for further offline processing and statistical analysis.
\vspace{-0.25cm}
\section{Measurement Process and Loading of Enclosure}
\begin{figure}
    \centering
    \includegraphics[width=\columnwidth]{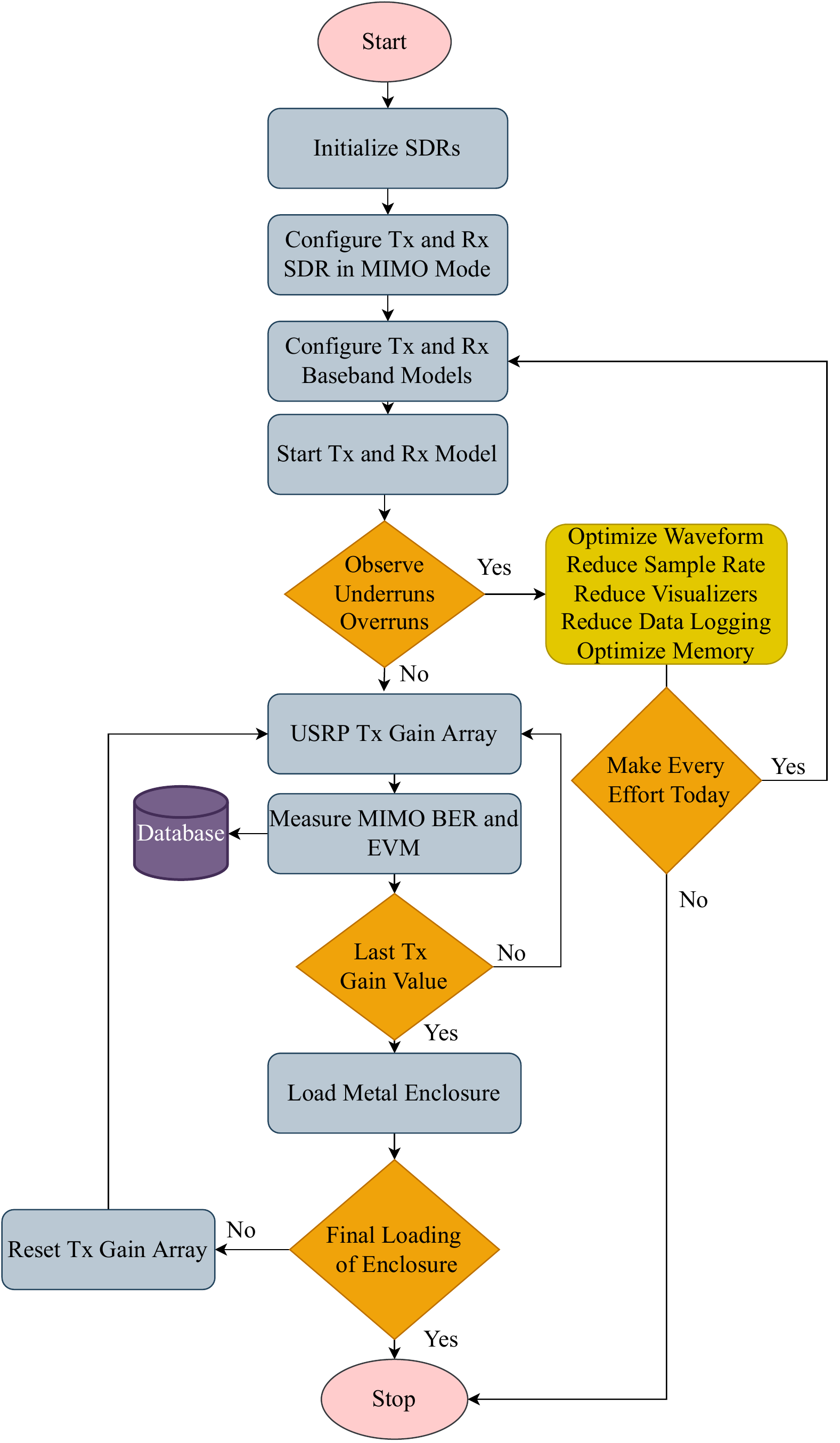}
    \caption{Flowchart of MIMO BER and EVM measurement.}
    \label{fig:flowchart_ber_evm}
\end{figure}

We used transmitter and receiver baseband models with QPSK modulation for over-the-air-transmission. After the Tx and Rx USRPs are initialized and configured in MIMO mode, the Tx and Rx QPSK baseband models are also configured to operate in MIMO communication. Baseband transmitter generates two-column QPSK modulated waveforms that are passed to X310 USRP for over-the-air transmission in metal enclosure. Two-column QPSK modulated waveforms represent each channel of the USRP. Digital samples from host PC to USRP for upconversion and transmission are transferred using 1Gig Ethernet cable. Another X310 USRP captures the received signals which are digital down-converted and passed to host PC for baseband processing and two channel BER and EVM measurements. Measurements are conditioned only to proceed when there are no underruns and overruns respectively. Underruns and overruns affect the reliability of KPI measurements particularly two-channel BER measurements. Underruns and overruns depends on the base sampling rates of the USRP, waveform design, speed of the cable connecting USRP and host PC, simulation time and the data-logging techniques in baseband software. X310 USRP can offer higher sampling rates to stream the data to and from host PC, but in most of the cases sample rates limited by the speed of Ethernet cable or the general purpose processors are not configured to handle maximum sampling rates of the USRPs. The Simulink Tx and Rx baseband models were configured to stop in case of any single underrun and/or overruns event occurs. USRP Tx gain is configured in Simulink and programmatically controlled so that two-channel near-field MIMO BER and EVM measurements are performed without lifting the metal lid off and creating any possibility of changing EM environment and propagation conditions. The propagation environment can be changed after the BER and EVM measurements are performed. The flowchart of the measurement process is shown in Fig. \ref{fig:flowchart_ber_evm} which shows the critical design and measurement steps.

\vspace{-0.25cm}
\section{Experimental Results and Discussions}
In this section, we present near-field MIMO BER and EVM measurement results in different scenarios. Measurements are performed in an empty enclosure in the presence of a static mode-stirrer and the near-field MIMO BER and EVM results are presented when the mode-stirrer is moving continuously. In the third scenario, we present near-field MIMO BER and EVM measurements in the presence of MIMO interference. MIMO interference is continuously generated using B210 USRP at sampling rates of 400 ksps. Baseband interference signals were generated in GNU Radio and passed to B210 USRP for the repeated transmission.
\subsection{Near-field MIMO BER and EVM in Empty Enclosure}
In this scenario 2x2 MIMO setup is investigated for BER and EVM measurements in stationary channel conditions. This scenario has direct coupling and as well as rich scattering.
\begin{table}
\centering
  \caption{2x2 near-field MIMO BER and EVM measurement in the metal enclosure. Maximum power gain of AGC is set to 60 and RF frequency is 5.6 GHz.}
  \begin{tabular}{|c|l|l|l|l|l|l|}
    \hline
    \multirow{3}{*}{\thead{Tx Gain\\ (dB)}} &
      \multicolumn{3}{c|}{Channel 1} &
      \multicolumn{3}{c|}{Channel 2} \\
   \cline{2-7} & Prx(dB)&\thead{EVM\\ ($\%$)}&BER&Prx(dB)&\thead{EVM\\ ($\%$)}&BER \\
    \hline
    2 & -71 & 87& 3e-2& -68& 81 &7e-3\\
    4&-69&84&7e-3&-66&77&3e-3\\ 
    6&-67&79&2e-3&-64&70&2e-3\\ 
    8&-61&58&3e-3&-61&61&4e-3\\
    \hline
  \end{tabular}
  \label{tab:ber_evm_60}
\end{table}
Table. \ref{tab:ber_evm_60} shows BER and EVM measurements of two USRP channels at frequency of 5.6 GHz and AGC gain of 60. It can be seen that the BER and EVM of two USRP channels improves as the TX gain of the USRP is increased.
\begin{table}
\centering
 \caption{2x2 near-field MIMO BER and EVM measurement in the metal enclosure. Maximum power gain of AGC is set to 80 and RF frequency is 5.6 GHz.}
  \begin{tabular}{|c|l|l|l|l|l|l|}
    \hline
    \multirow{3}{*}{\thead{Tx Gain \\(dB)}} &
      \multicolumn{3}{c|}{Channel 1} &
      \multicolumn{3}{c|}{Channel 2} \\
   \cline{2-7} & Prx($\mathrm{dB}$)&\thead{EVM\\ ($\%$)}&BER&Prx(dB)&\thead{EVM \\($\%$)}&BER \\
    \hline
   2&-67&22&3.4e-2&-68&20&5.3e-3\\
   4&-65&17&5.1e-3&-66&18&4.7e-3\\
   6&-63&14&9.3e-3&-64&15&9.3e-3\\
   8&-61&13&5.4e-3&-61&13&5.1e-3\\
    \hline
  \end{tabular}

 \label{tab:ber_evm_80}
\end{table}
Table \ref{tab:ber_evm_80} shows BER and EVM measurement of two USRP channels at a frequency of 5.6 GHz and AGC gain of 80. It can be seen that the BER and EVM of the two USRP channels improve as the Tx gain of the USRP is increased. A noticeable difference in the measured BER and EVM values can be observed for AGC gain of 60 and 80 respectively. For both the cases BER and EVM values improve however, the BER values decrease faster for AGC gain value of 80. This is because of the extra increase in the AGC gain.
\begin{figure}
    \centering
    \includegraphics[width=0.45\linewidth]{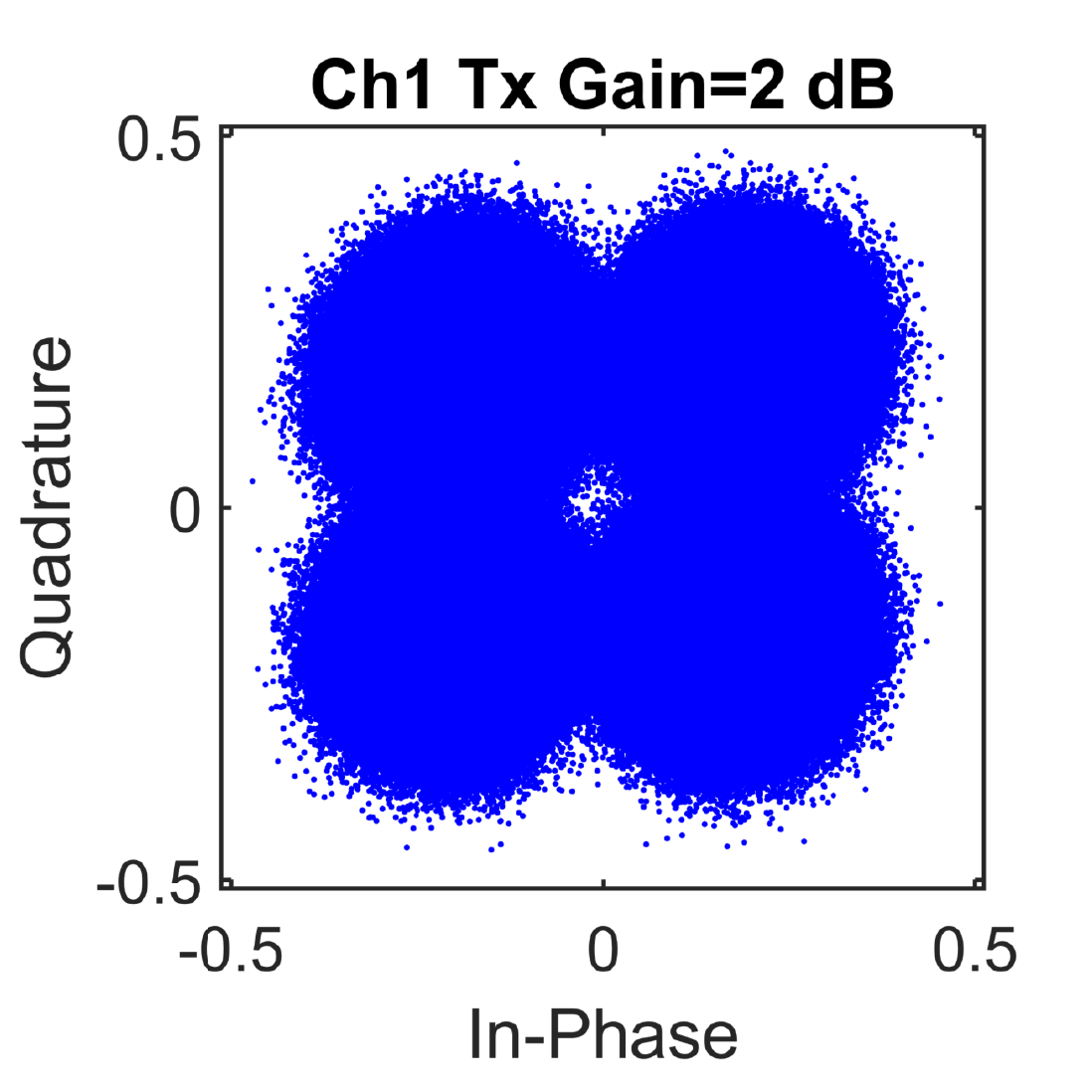}
    \includegraphics[width=0.45\linewidth]{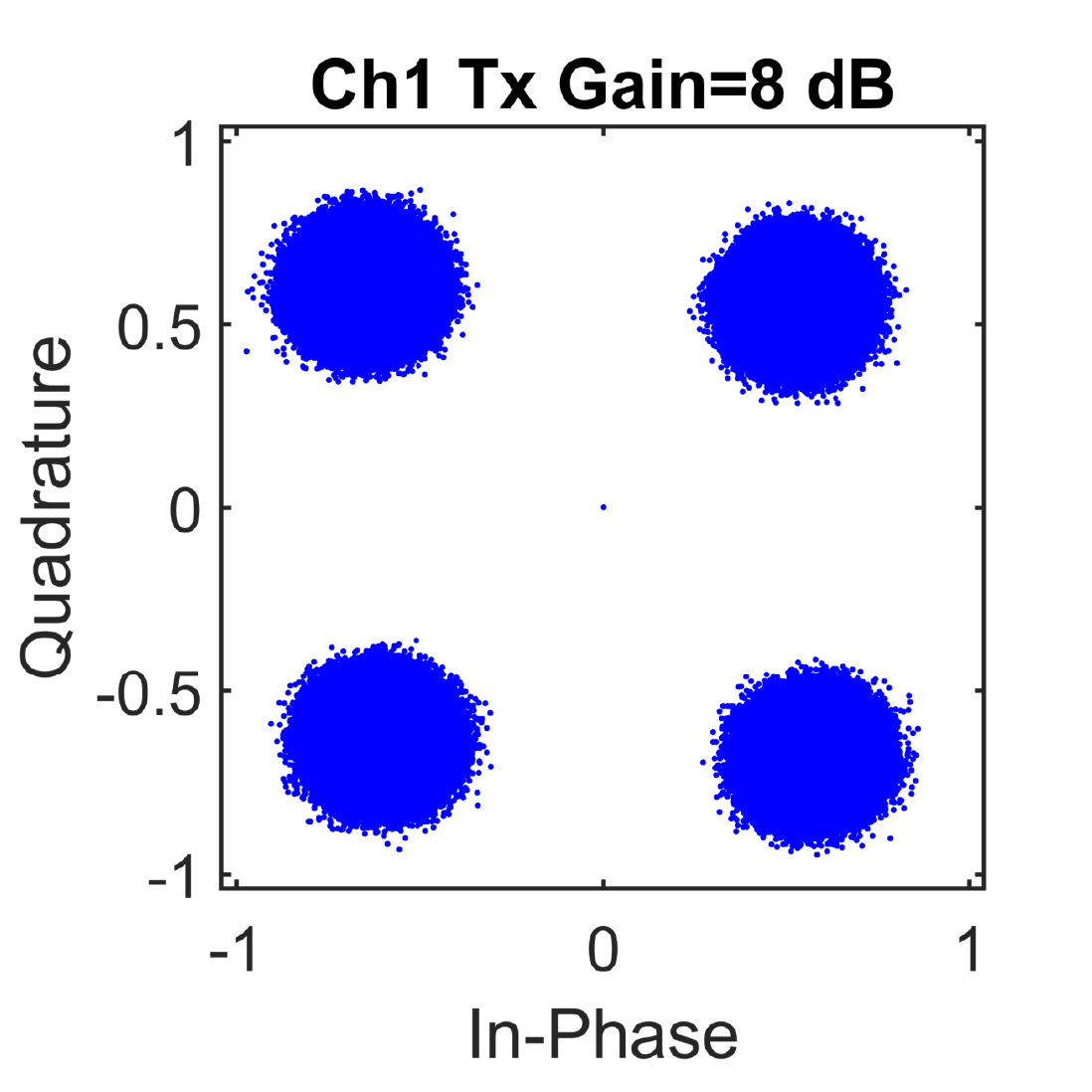}\\
    \includegraphics[width=0.45\linewidth]{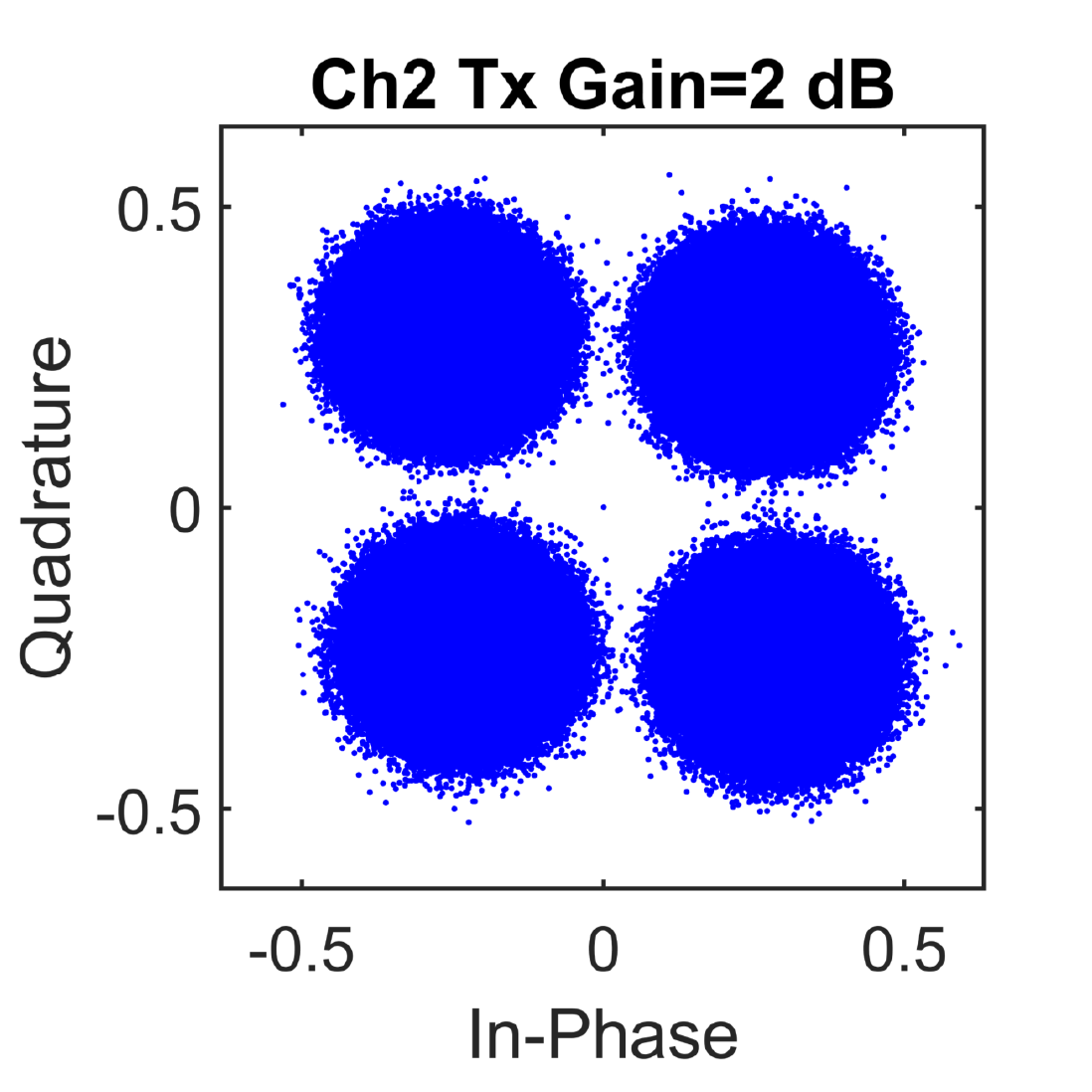}
    \includegraphics[width=0.45\linewidth]{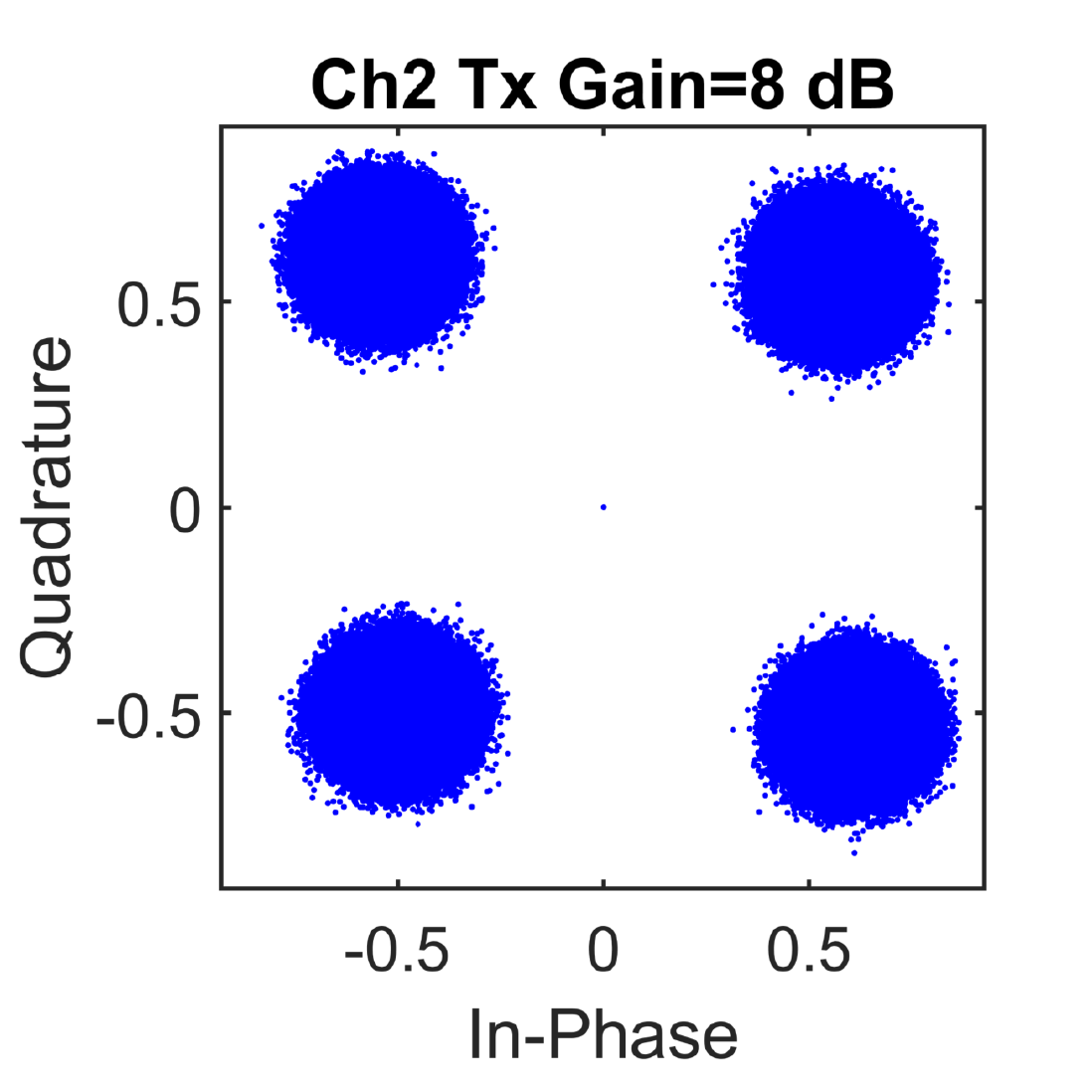}
    \caption{Channel 1 and channel 2 QPSK constellation diagram at 5.6 GHz. AGC gain=60, Rx Gain=0 dB and Tx Gain varied from 2 dB to 8 dB with step of 2 dB.}
    \label{fig:const_carrier_60}
\end{figure}
Fig, \ref{fig:const_carrier_60} shows QPSK constellation diagram of two USRP channels. Measured QPSK constellations shown in this work are recorded after carrier frequency synchronization. The QPSK constellation diagram is recorded after carrier synchronization for different Tx gain values of the USRPs. It can be seen that the USRP channel 2 is better than USRP channel 1. Additionally, the two USRP channel QPSK constellation diagram improves as the Tx gain of the USRP increases which corresponds to improved BER and EVM measurements.

Similar to the theoretical predictions and previous measurements, the two USRP channel QPSK constellation diagrams improve as the Tx gain of the USRP increases which corresponds to improved BER and EVM measurements. The measurements are performed at stationary channel conditions when the mode-stirrer was off. The mode-stirrer rotation has direct implications on the measured BER and EVM values because when the mode-stirrer rotates channel becomes non-stationary. The channel power gain of the MIMO communication link varies drastically as a result AGC at the receiver tries to compensate low and high gain values to stabilize the received signal for carrier frequency offset correction. Hence EVM measurement of two USRP channels is not stationary because of time-varying channel power gain and time-varying frequency offset. The same measurements were repeated when the mode-stirrer was turned on. In the presence of mode-stirrer rotation the MIMO channel is time-varying, therefore the constellation diagram and MIMO digital receiver KPIs are also time-varying.
\vspace{-0.15cm}
\subsection{EVM Measurements Moving Stirrer}
\begin{figure*}
    \centering
    \includegraphics[width=0.22\linewidth]{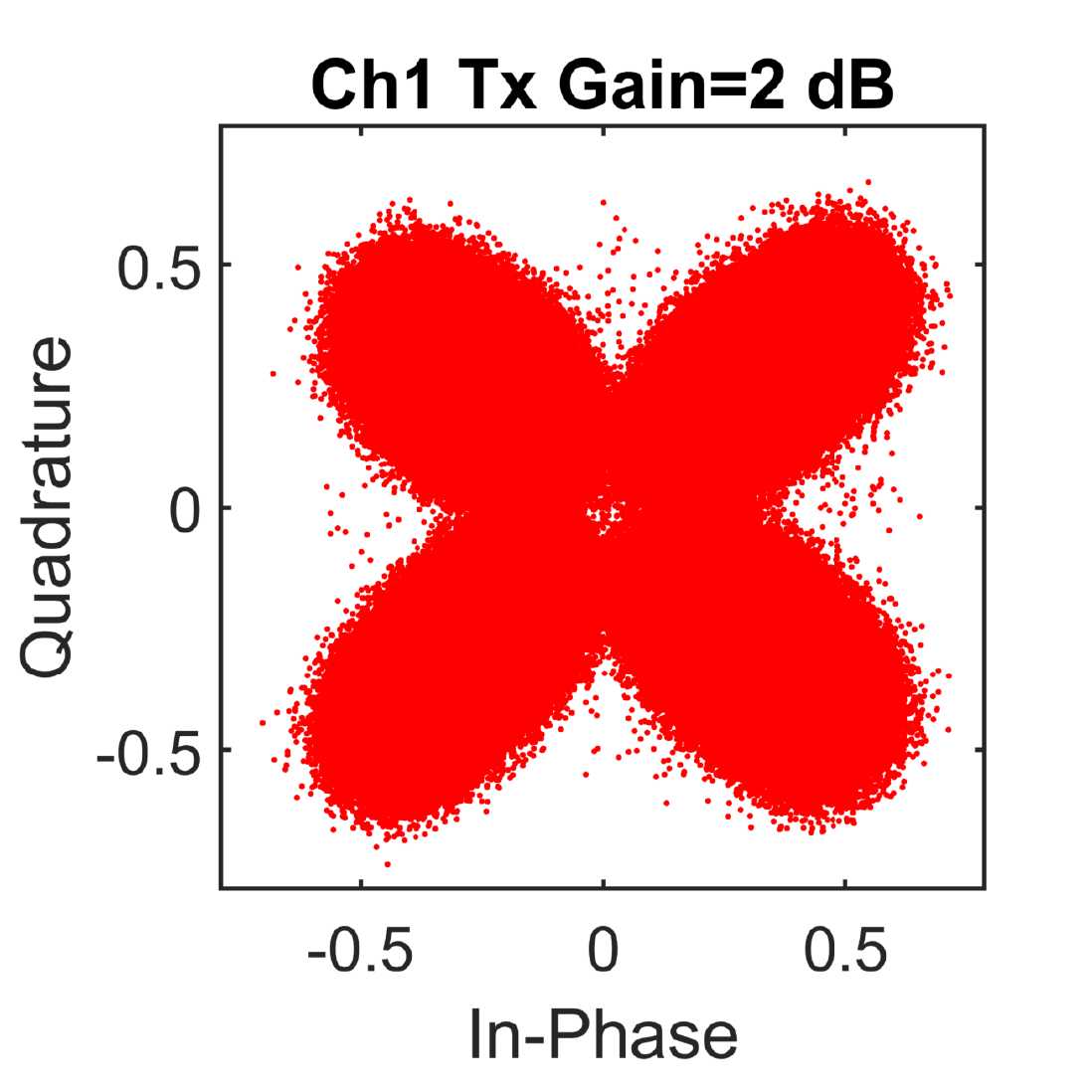}
    \includegraphics[width=0.22\linewidth]{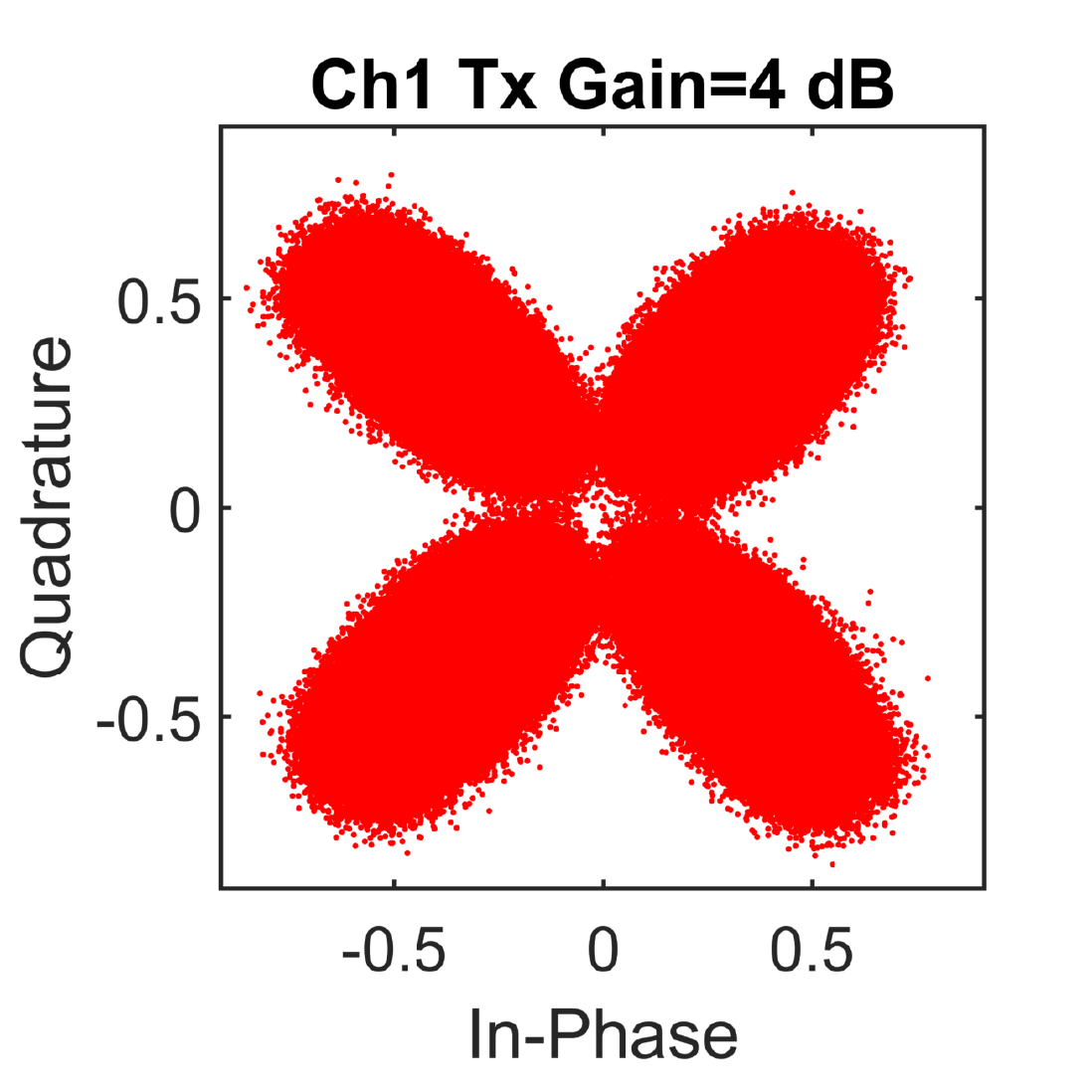}
    \includegraphics[width=0.22\linewidth]{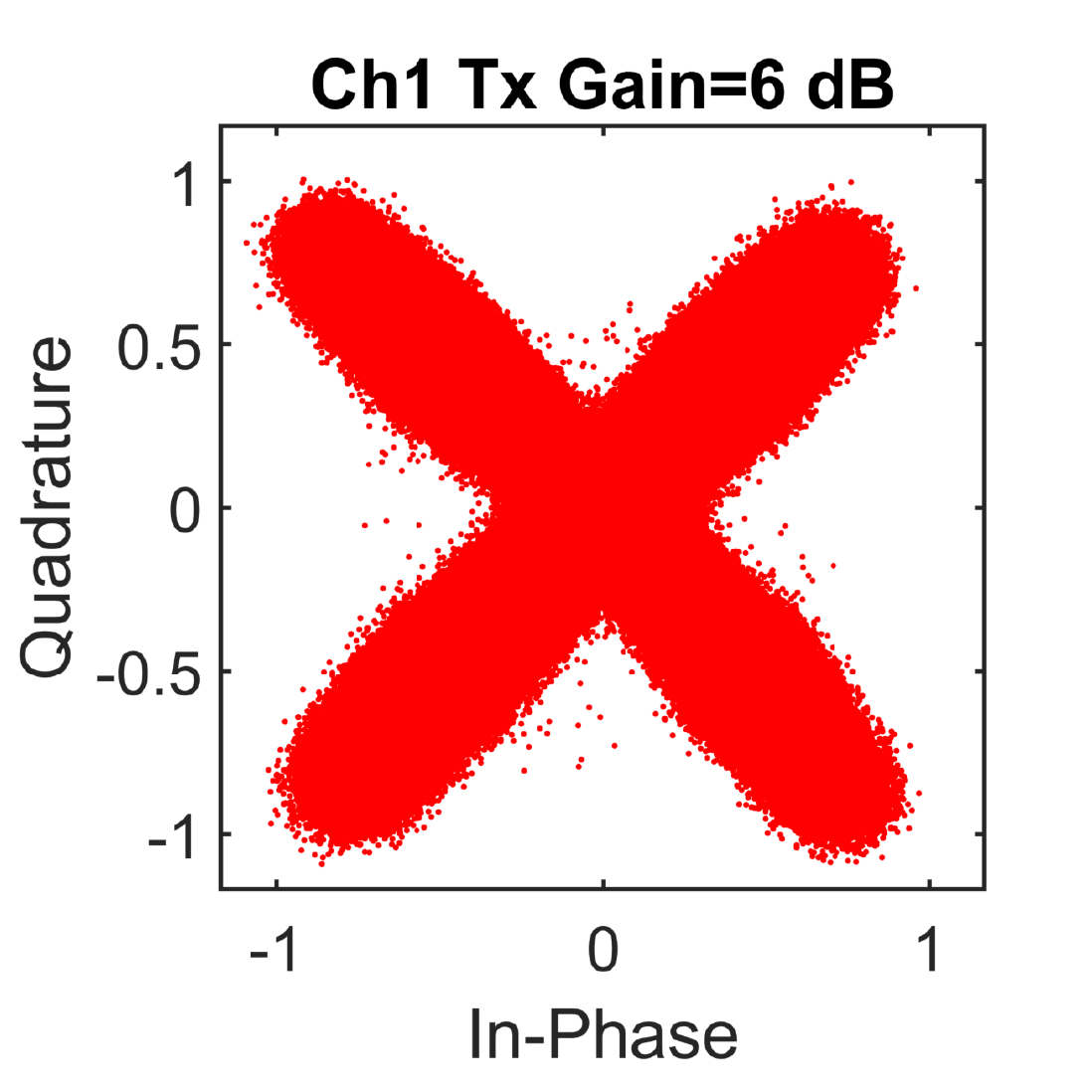}
    \includegraphics[width=0.22\linewidth]{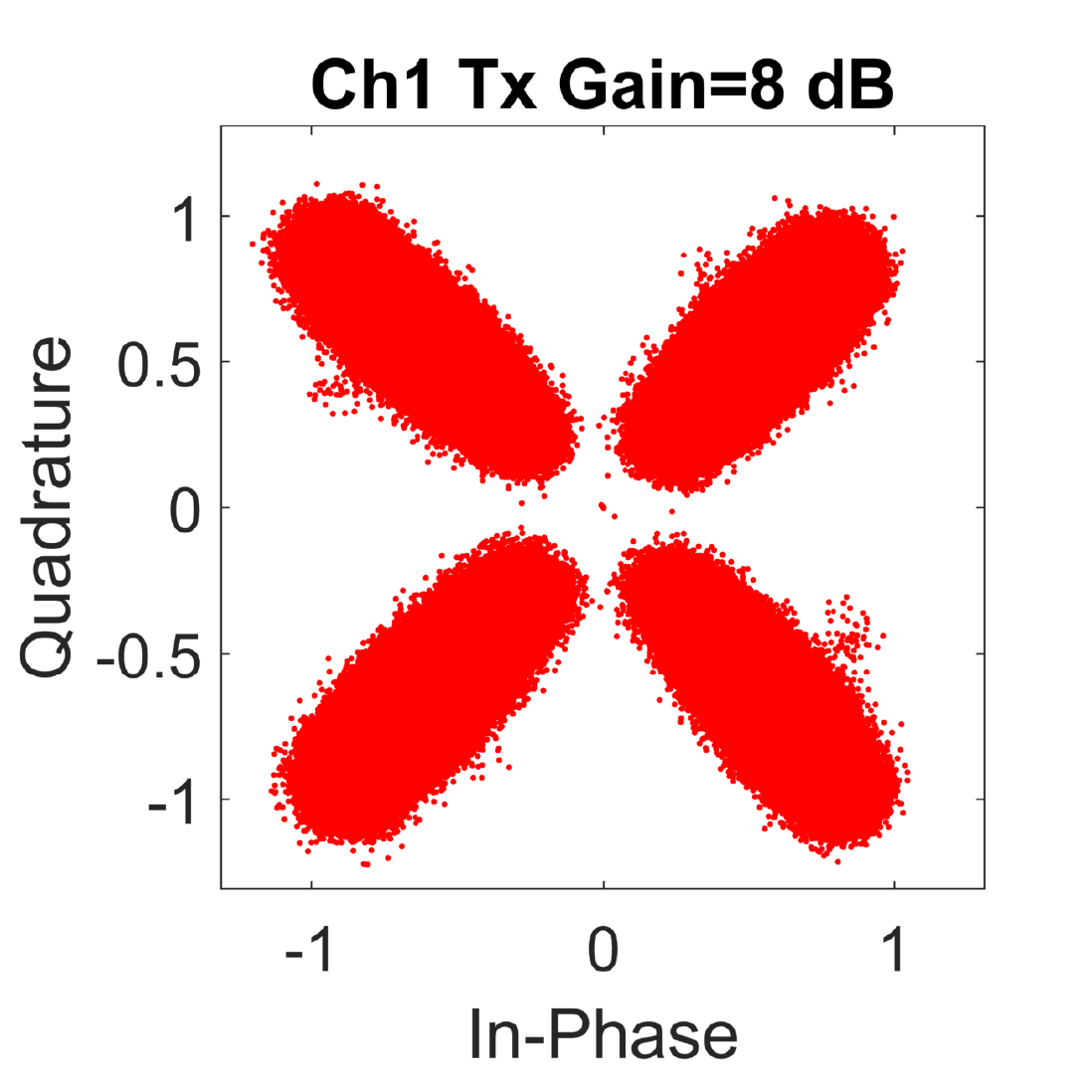}\\
    \includegraphics[width=0.22\linewidth]{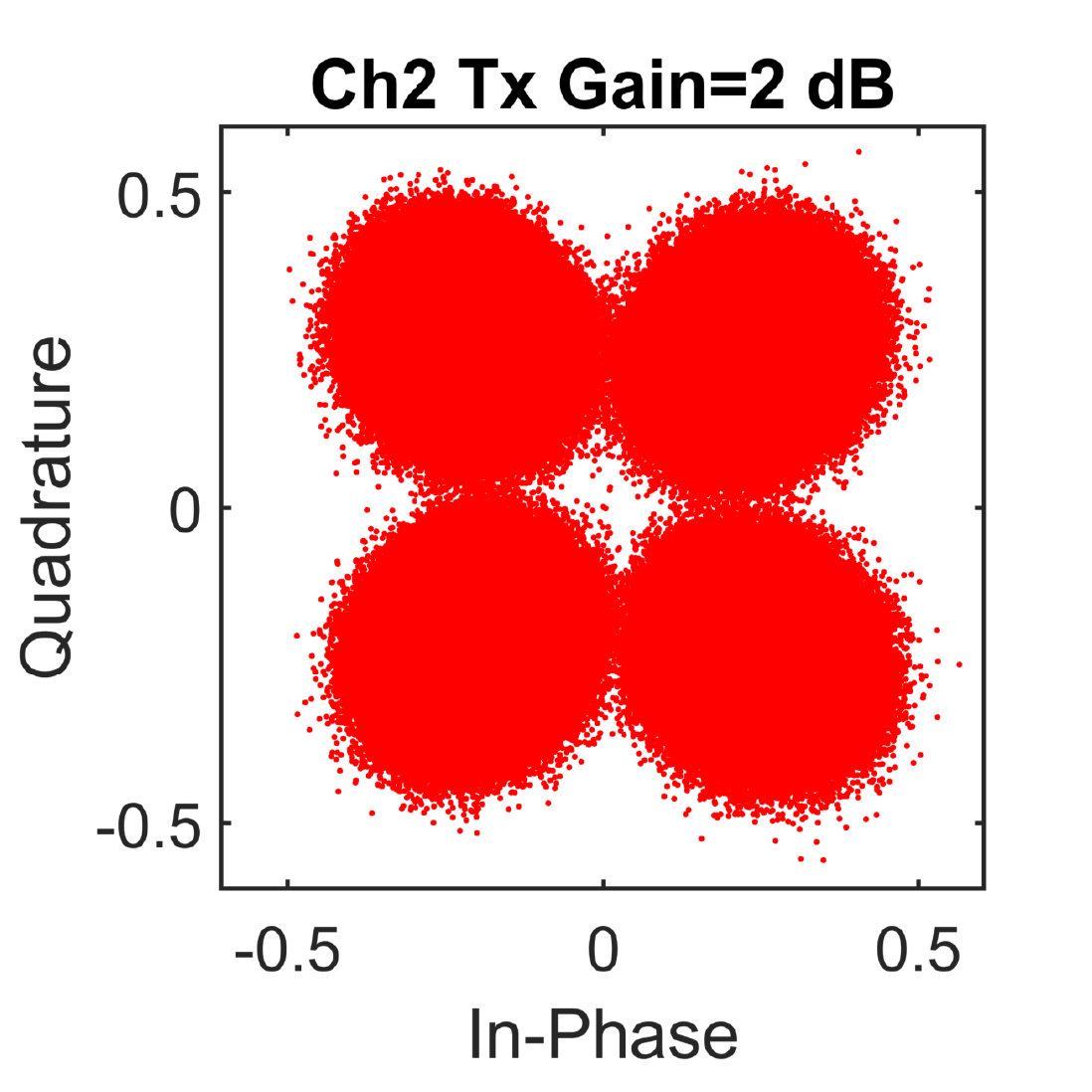}
    \includegraphics[width=0.22\linewidth]{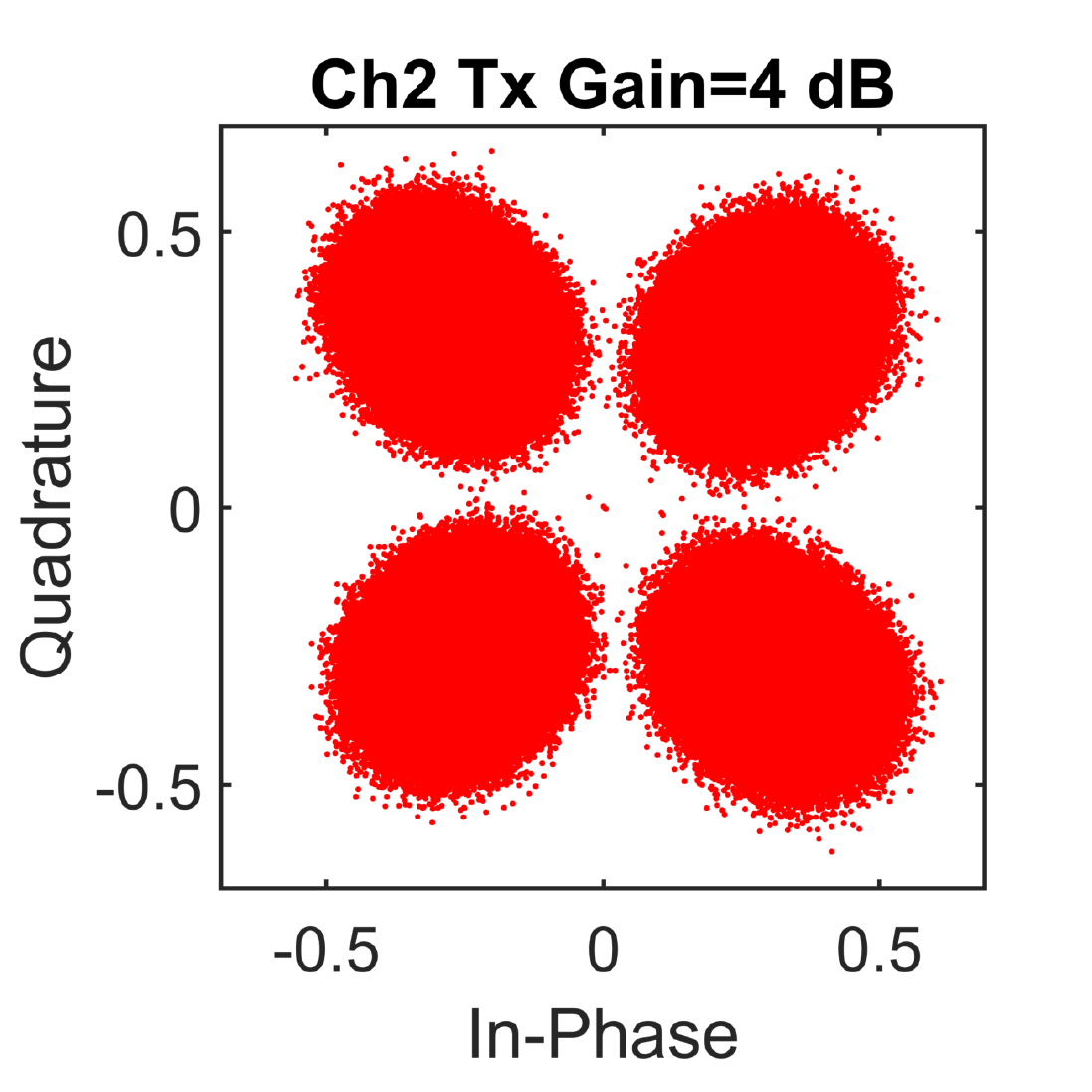}
    \includegraphics[width=0.22\linewidth]{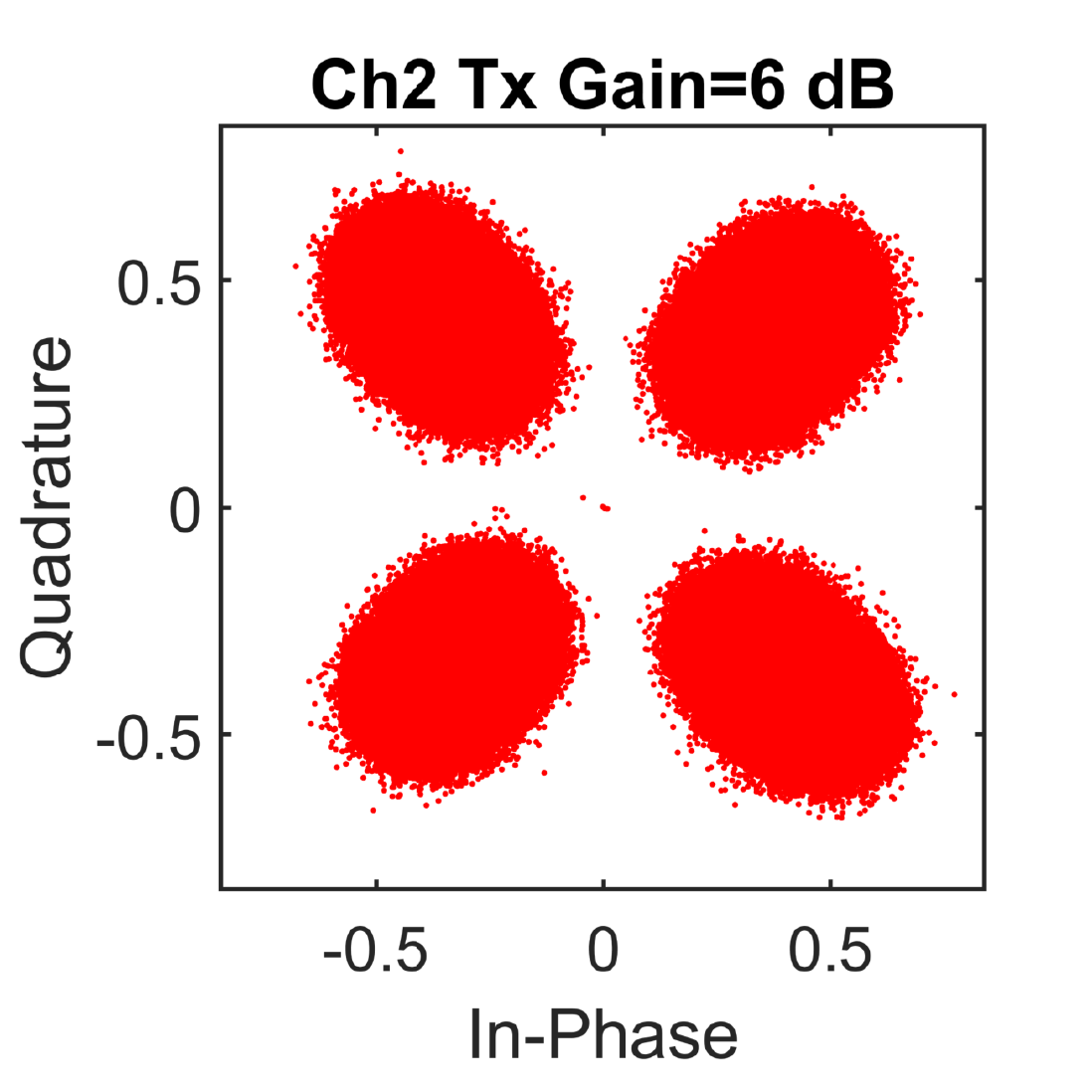}
    \includegraphics[width=0.22\linewidth]{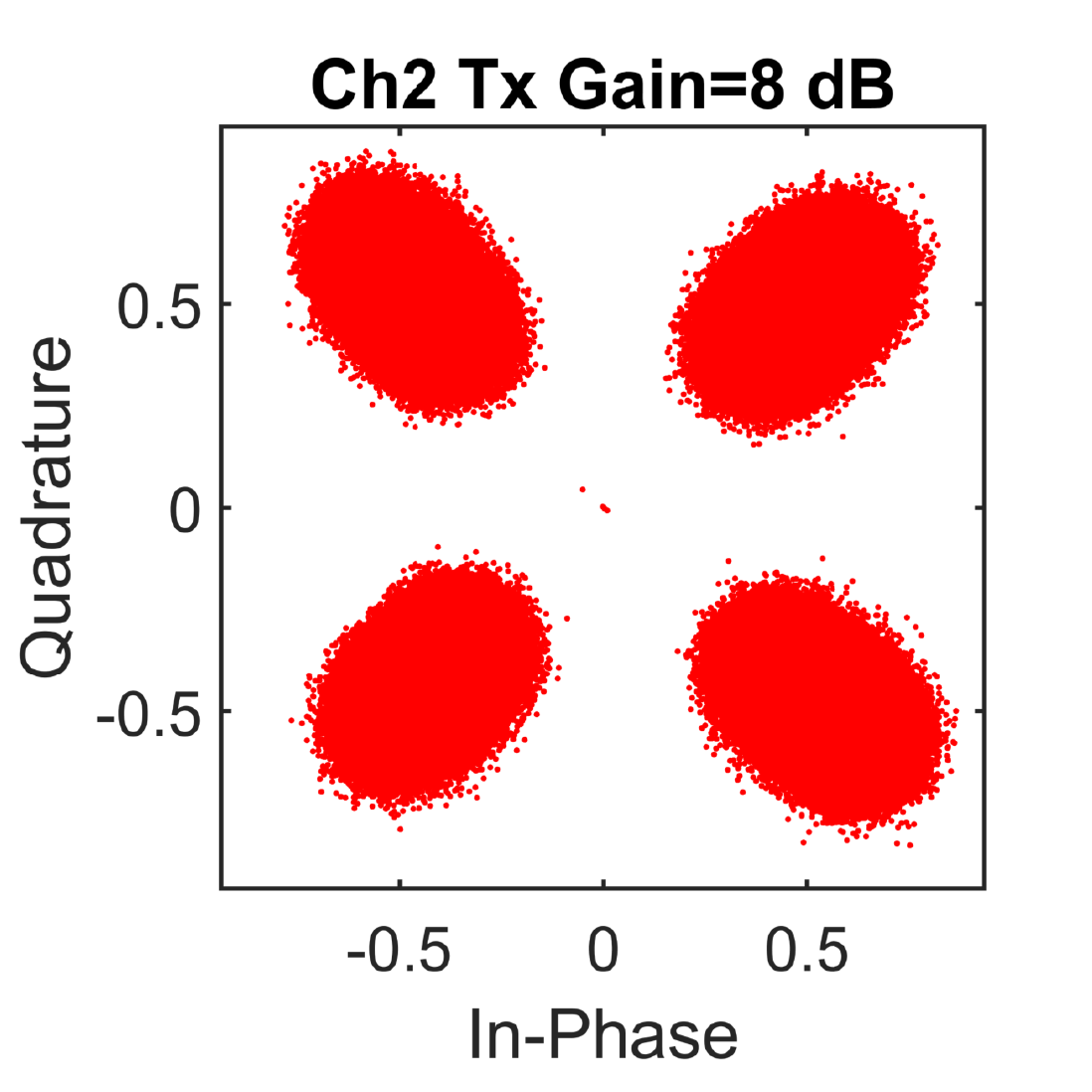}
    \caption{Channel 1 and channel 2 QPSK constellation diagram at 5.6GHz in mode-stirred metal enclosure. AGC gain=60 dB, Rx Gain=0 dB and Tx Gain varied from 2 dB to 8 dB with step of 2 dB.}
    \label{fig:const_carrier_60_stirr_on}
\end{figure*}

\begin{figure}
    \centering
    \subfloat[\label{fig:ch12_evm_2dB}]{\includegraphics[width=0.9\linewidth]{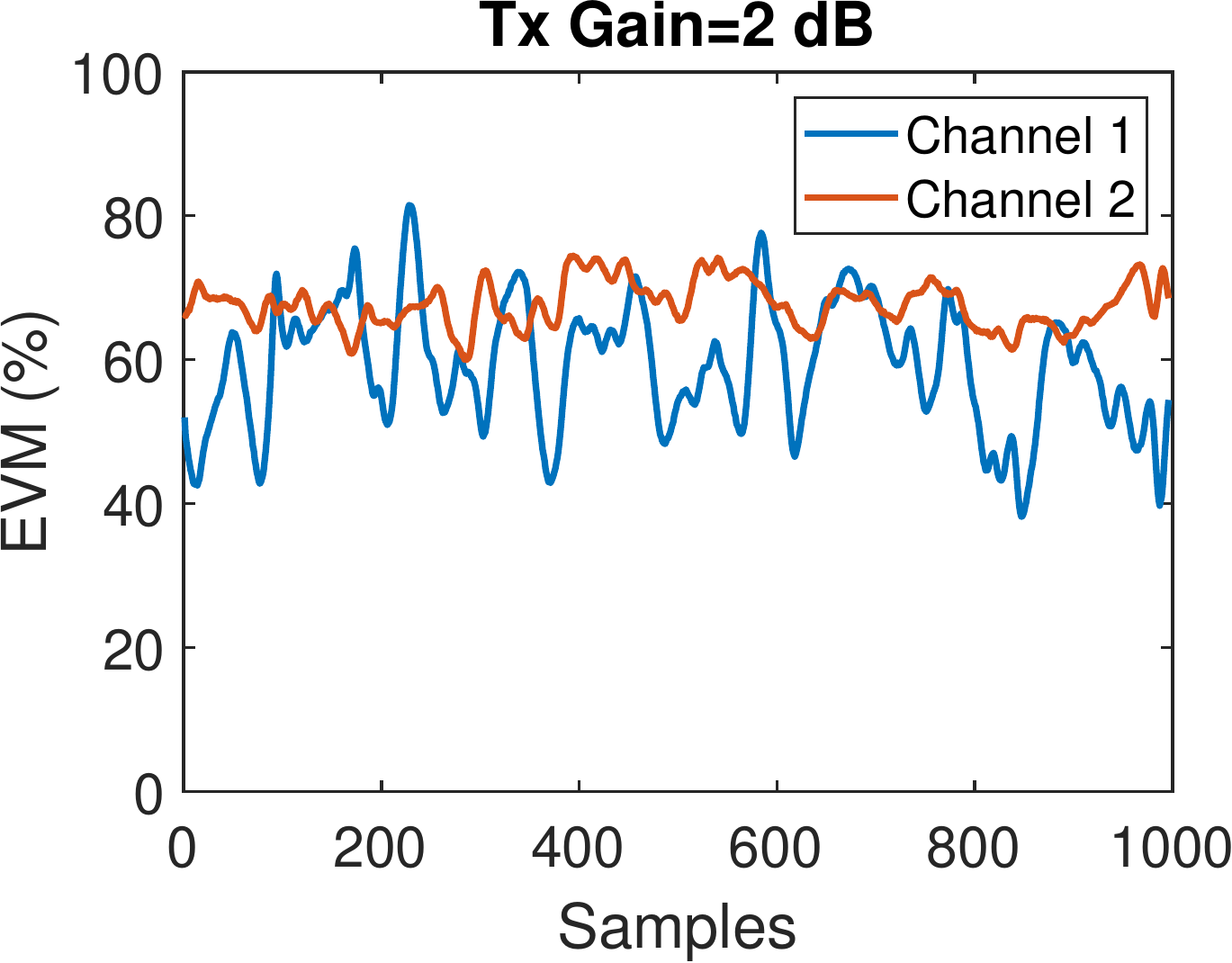}}\\
    \subfloat[\label{fig:ch12_evm_8dB}]{\includegraphics[width=0.9\linewidth]{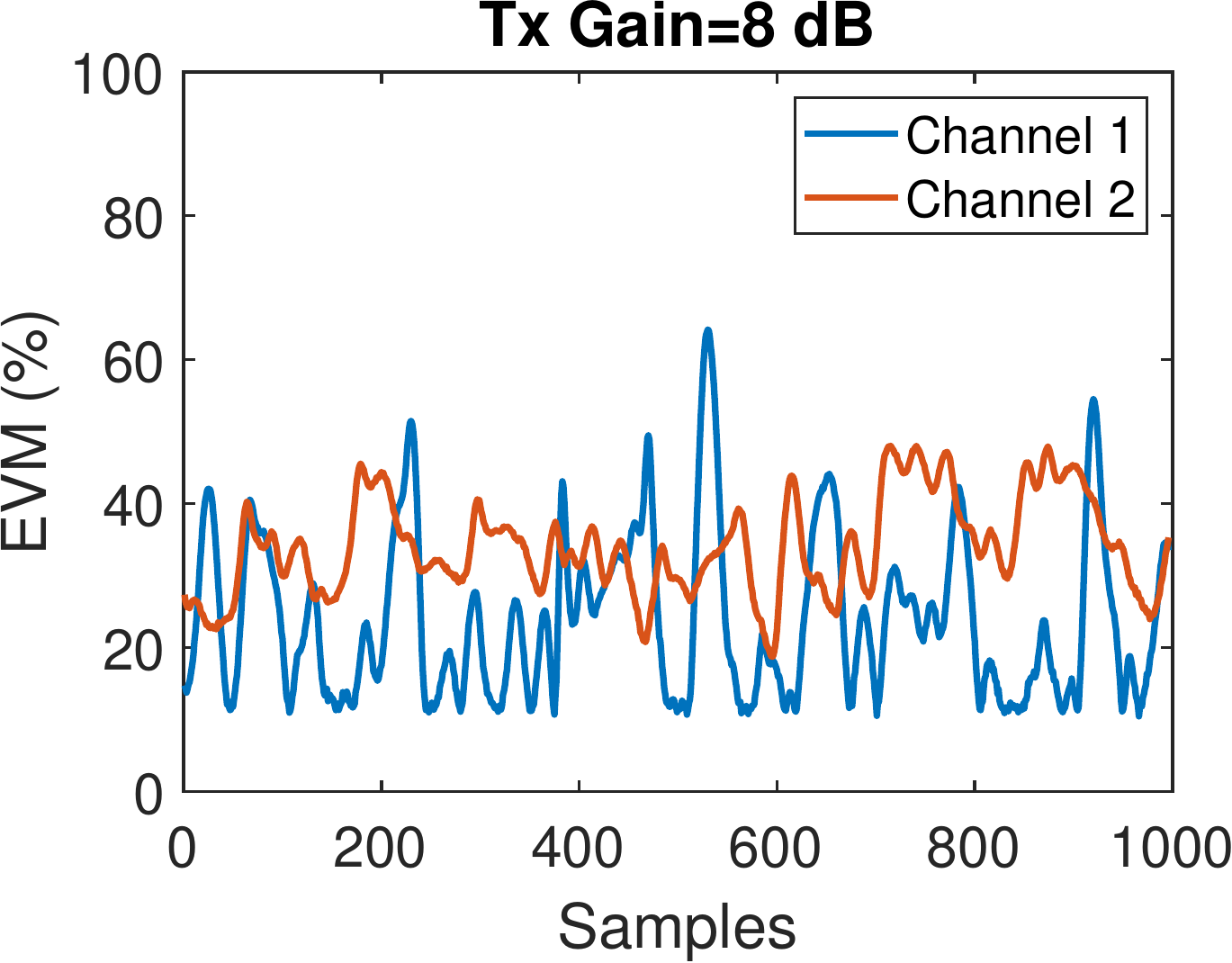}}
    \caption{Channel 1 and channel 2 instantaneous EVM measurement at 5.6GHz in mode-stirred metal enclosure. AGC gain=60 dB, Rx Gain=0 dB and Tx Gain varied from 2 dB to 8 dB with step of 2 dB.}
    \label{fig:ch12_EVM}
\end{figure}

EVM measurements of two USRP channel is measured when the stirrer is continuously moving. The movement of the stirrer creates a time-varying channel which has a direct implication on two-channel EVM. Figure \ref{fig:const_carrier_60_stirr_on} shows QPSK constellation diagram of two USRP channels. It can be seen that the QPSK constellation diagrams are also time-varying as the mode-stirrer rotates the received power values changes whose effect is visualized in the form of constellation diagram with low and high gain values. This means the BER and EVM are also time-varying. Average BER and EVM of two USRP channels is summarized in table \ref{tab:ber_evm_time_varying}. Fig. \ref{fig:ch12_EVM} shows time-varying EVM of two USRP channels for Tx gain values from 2 dB to 8 dB respectively. It can be seen that the USRP channel 1 EVM has a high variation in max and min values of EVM. USRP channel 2 is more stable. For the Tx gain value of 2 dB (see Fig.\ref{fig:ch12_evm_2dB}), the standard deviation of channel 1 EVM is $\sigma_{ch1}=1.93$ and the standard of channel 2 EVM is $\sigma_{ch2}=1.01$.
Similarly, for the Tx gain of 8 dB(see Fig. \ref{fig:ch12_evm_8dB}), the standard deviation of channel 1 is $\sigma_{1}=1.11$ and the standard deviation of channel 2 is $\sigma_{2}=0.414$. The standard deviation of channel 1 is higher for all the Tx gain values. Note that the possible cable faults were ruled out by exchanging the two USRP channels which resulted in exactly similar performance. In other words, USRP channel 1 EVM was better than USRP channel 2 EVM. In this way we precisely restricted our two-channel EVM and BER values to be a function of propagation environment.

\begin{table}
\centering
 \caption{Two Channel BER and EVM measurement in mode-stirred metal enclosure at 5.6GHz. Maximum AGC=60, Rx Gain=0 dB and Tx Gain is varied from 2 dB to 8 dB with step of 2 dB.}
  \begin{tabular}{|c|l|l|l|l|l|l|}
    \hline
    \multirow{4}{*}{\thead{Tx Gain \\(dB)}} &
      \multicolumn{3}{c|}{Channel 1} &
      \multicolumn{3}{c|}{Channel 2} \\
   \cline{2-7} & \thead{Avg.\\ Prx(dB)}&\thead{Avg.\\EVM \\($\%$)}&BER&\thead{Avg.\\ Prx(dB)}&\thead{Avg.\\ EVM\\($\%$)}&BER \\
    \hline
   2&-69&59&1.4e-2&-69&67&5.2e-3\\
   4&-65&49&7.3e-3&-67&59&8.3e-4\\
   6&-63&37&5.7e-3&-65&48&0\\
   8&-61&23&1.6e-3&-63&33&0\\
    \hline
  \end{tabular}
 \label{tab:ber_evm_time_varying}
\end{table}
\subsection{EVM Measurements in the presence of Interference}
\begin{figure}
    \centering
    \subfloat[]{\includegraphics[width=0.7\columnwidth]{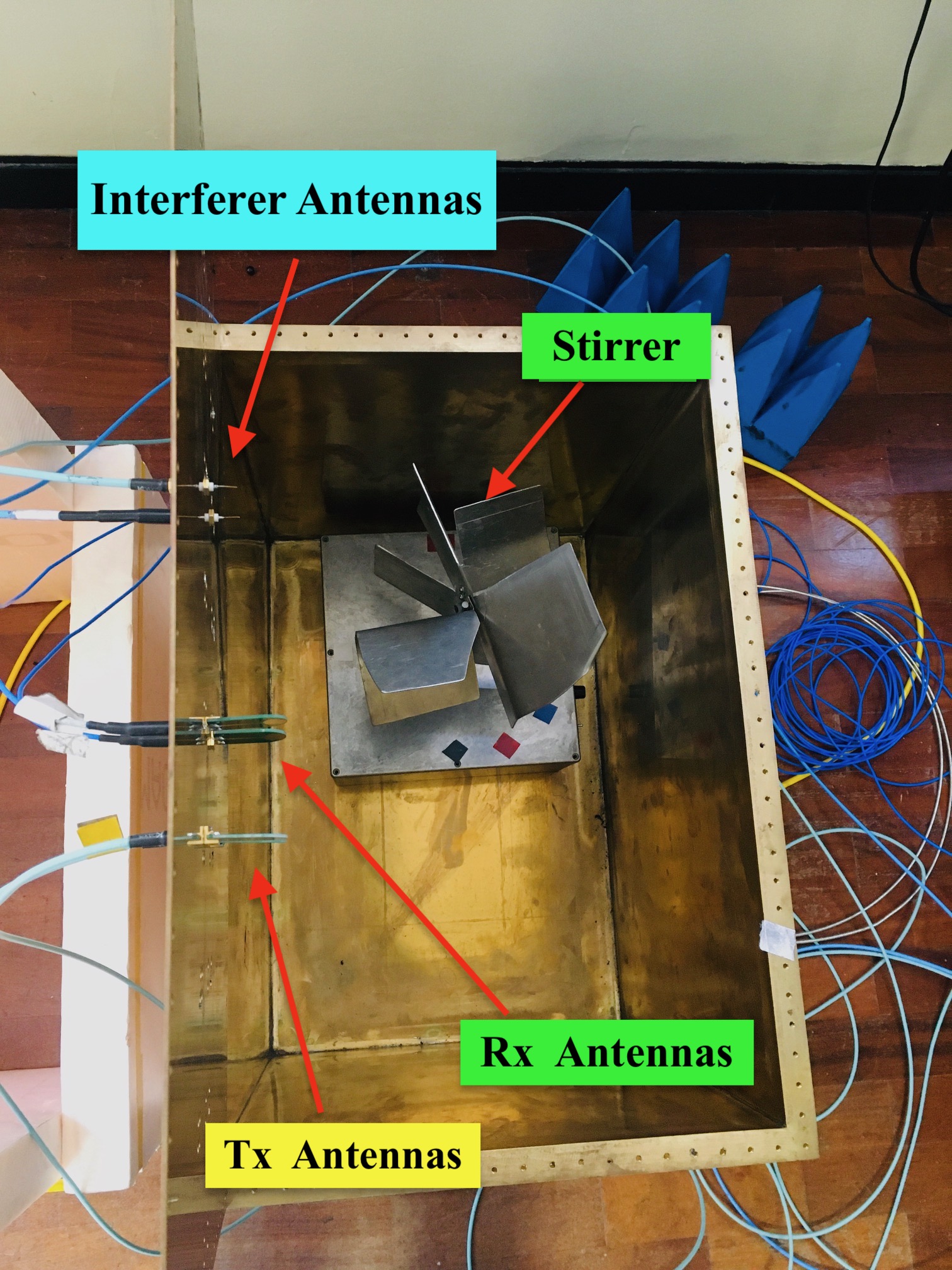}}\\
    \subfloat[]{\includegraphics[width=0.7\columnwidth]{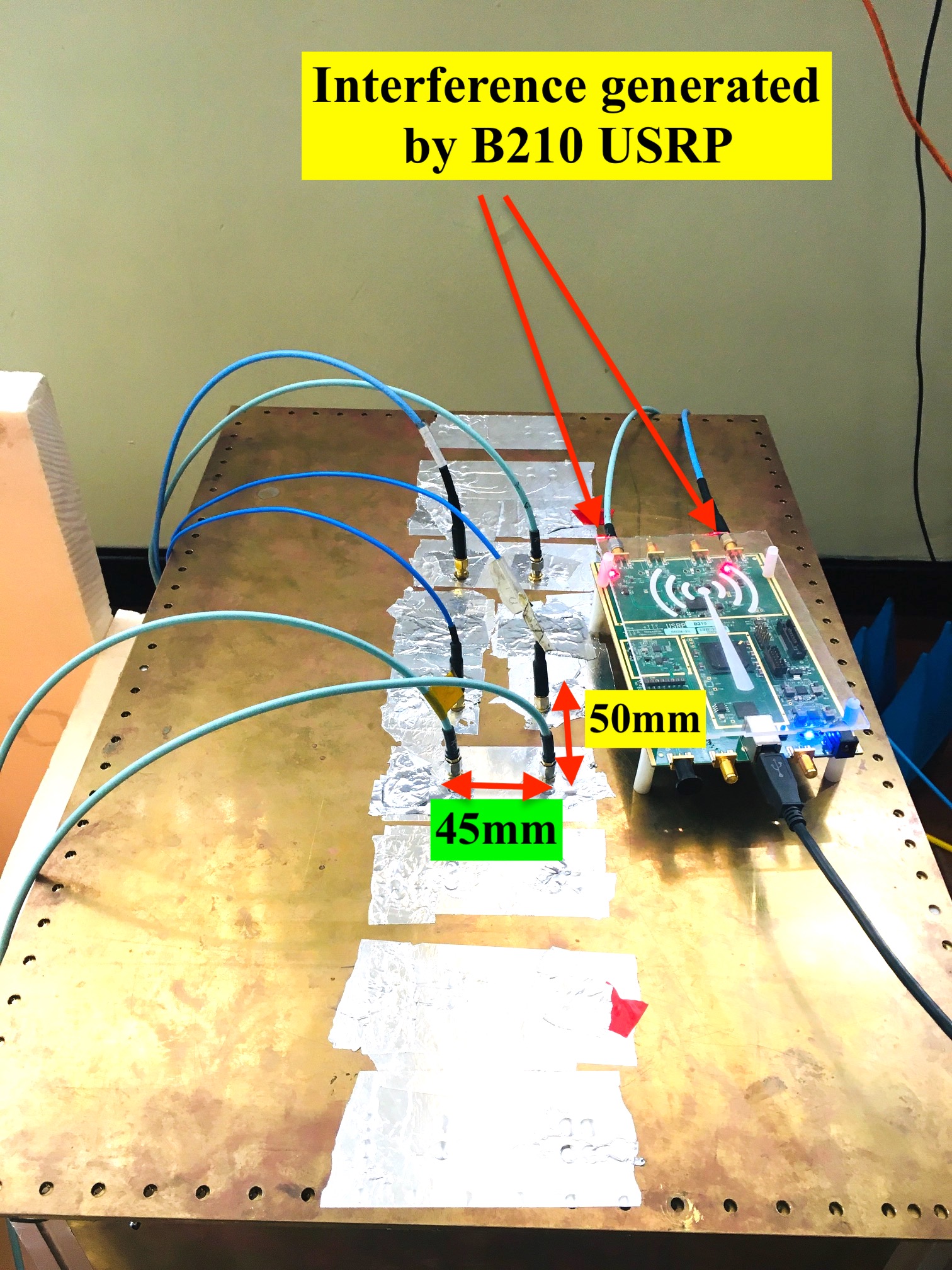}}
    \caption{2x2 MIMO BER and EVM measurement setup in the presence of MIMO interference. }
    \label{fig:topview_panel_int}
\end{figure}

\begin{figure}
    \centering
    \includegraphics[width=0.45\linewidth]{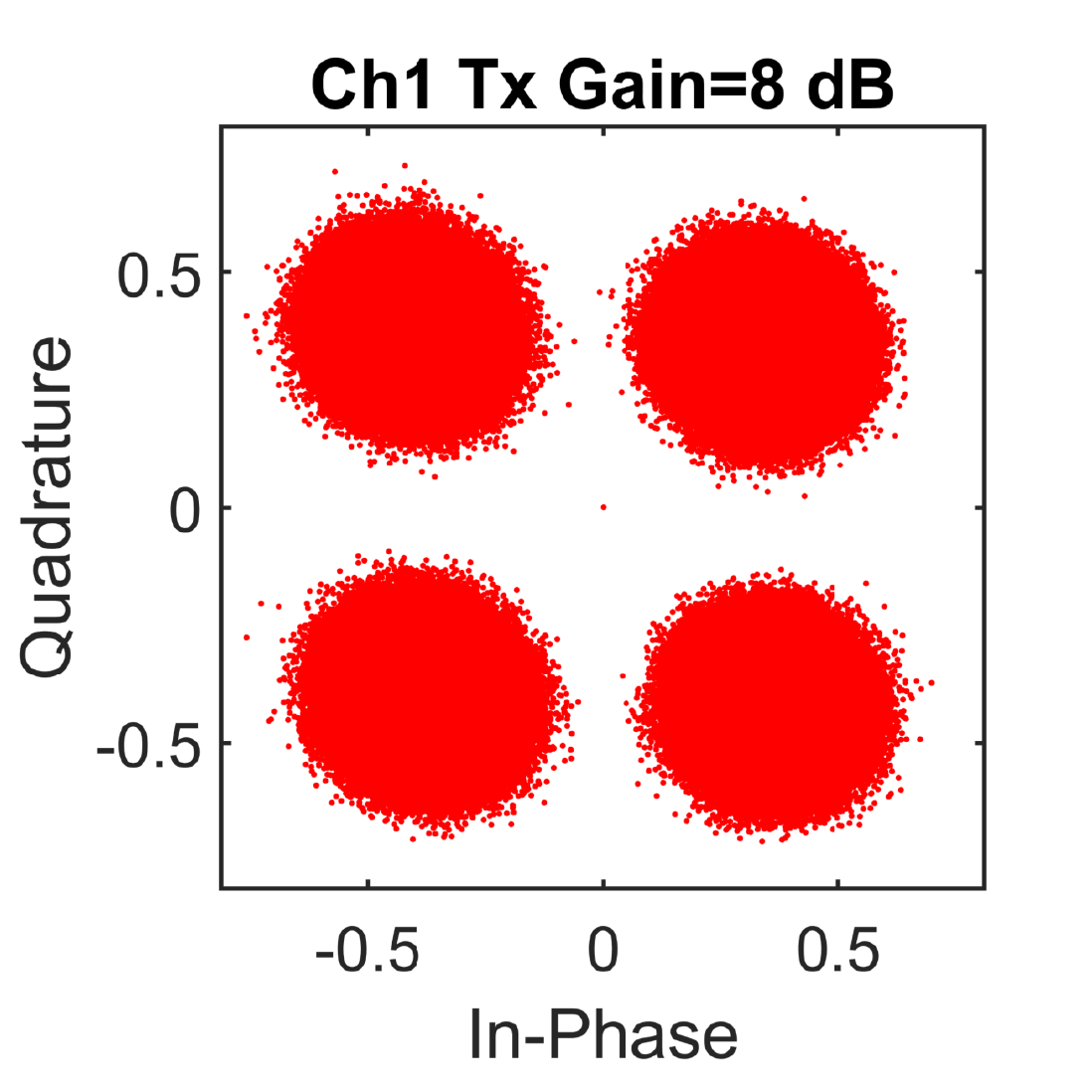}
    \includegraphics[width=0.45\linewidth]{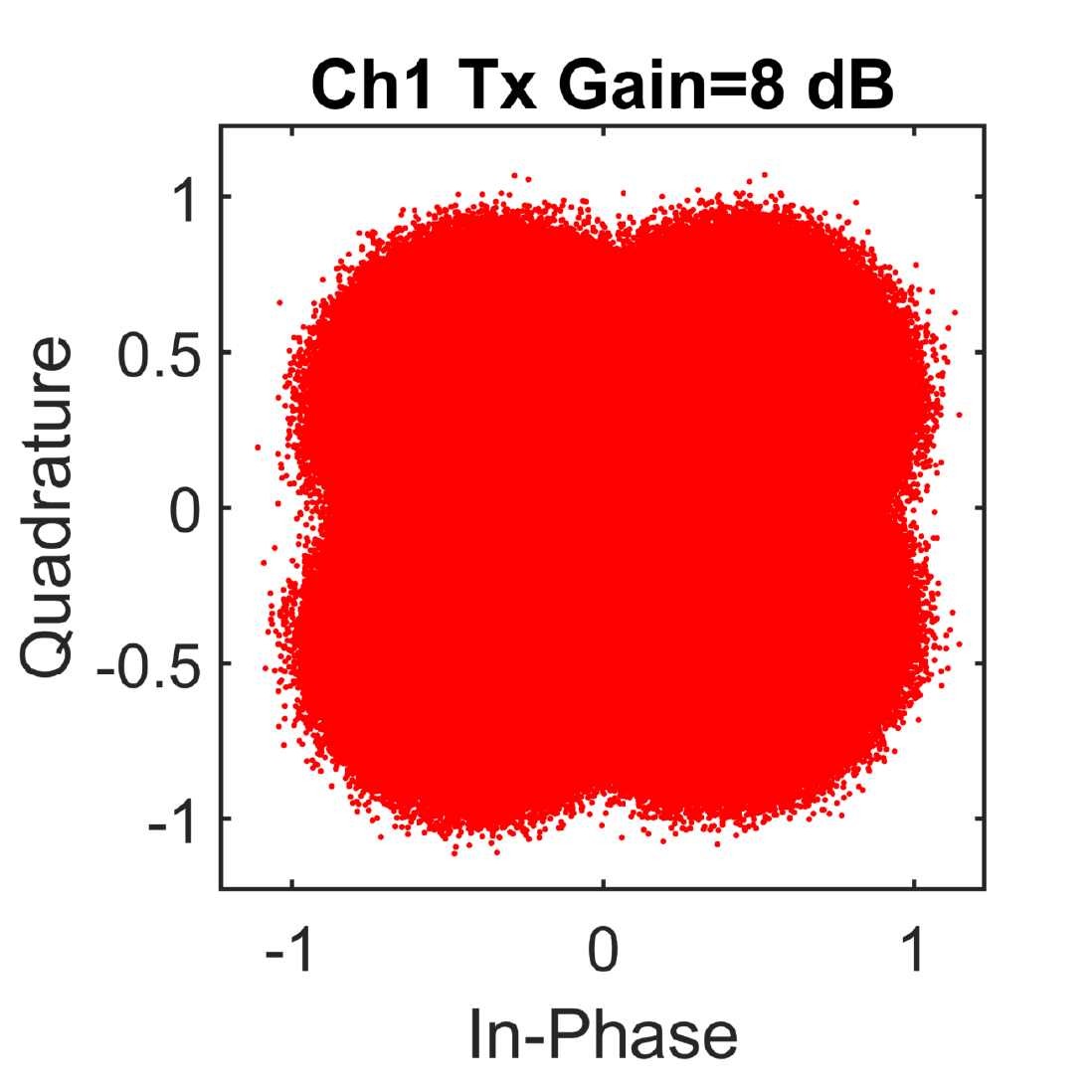}\\
    \includegraphics[width=0.45\linewidth]{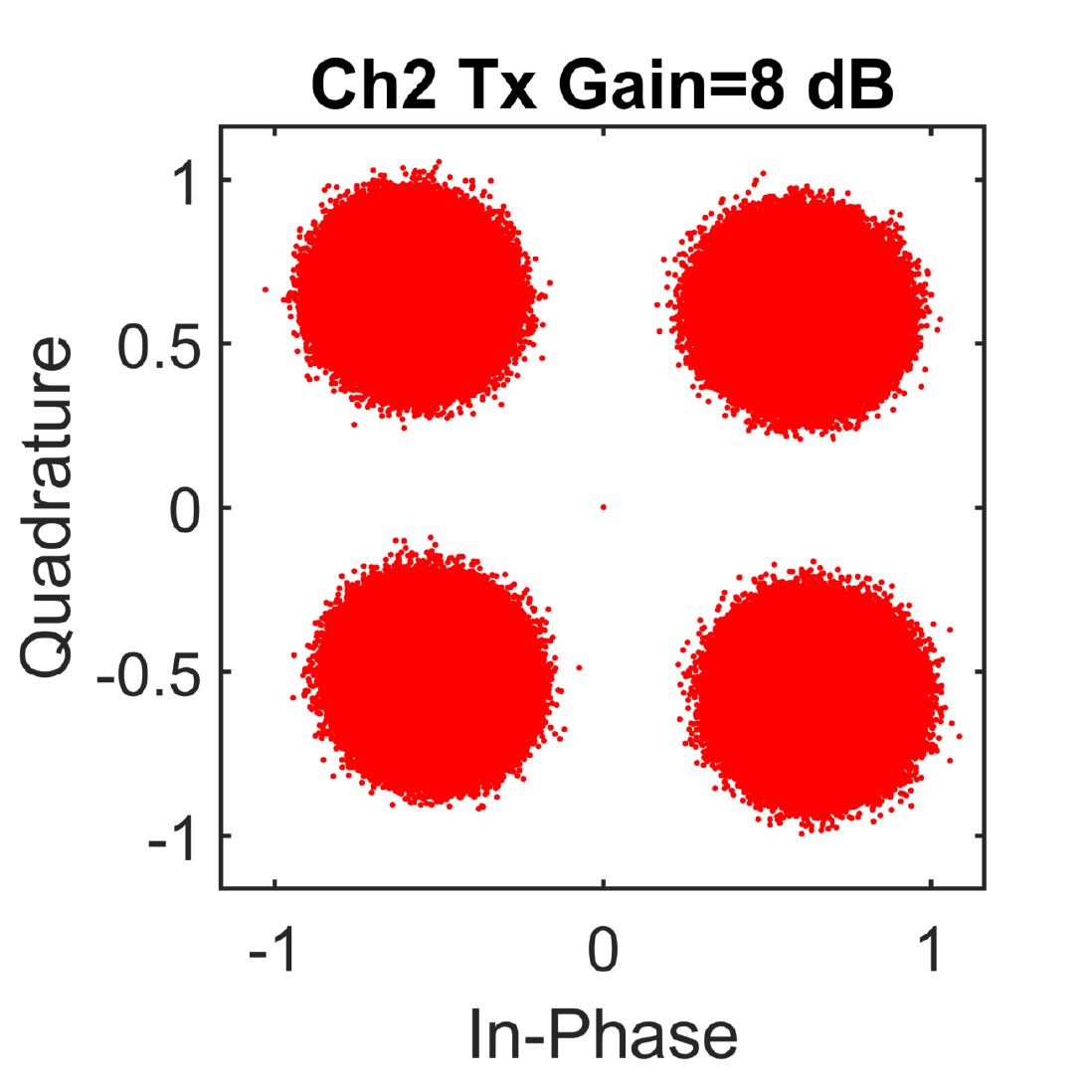}
    \includegraphics[width=0.45\linewidth]{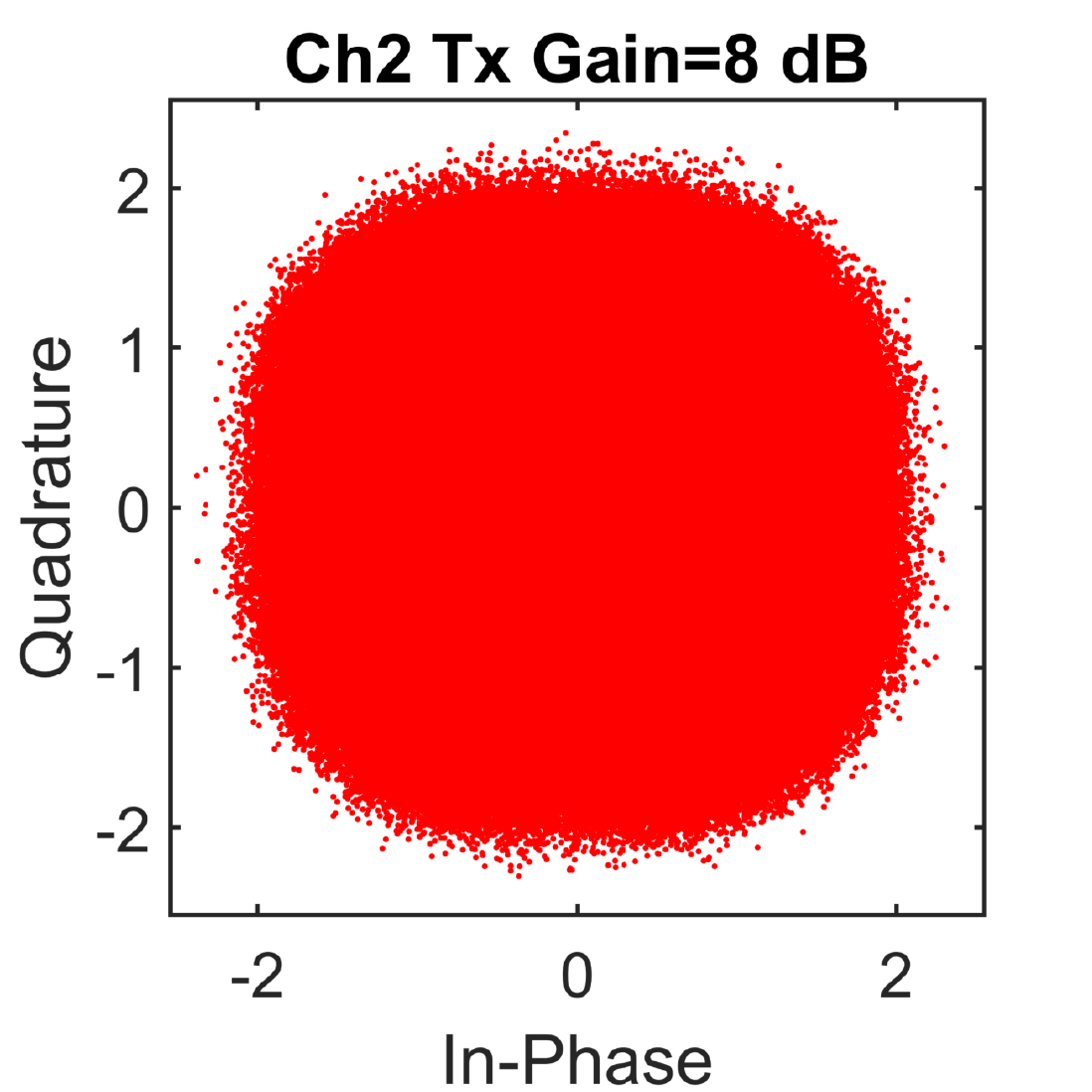}
    \caption{Channel 1 and channel 2 QPSK constellation diagram at 5.6GHz in the metal enclosure in the presence of MIMO interference. AGC gain=60 dB, Rx Gain=0 dB and Tx Gain= 8 dB.}
    \label{fig:const_qpsk_60_int}
\end{figure}

\begin{figure}
    \centering
    \includegraphics[width=\columnwidth]{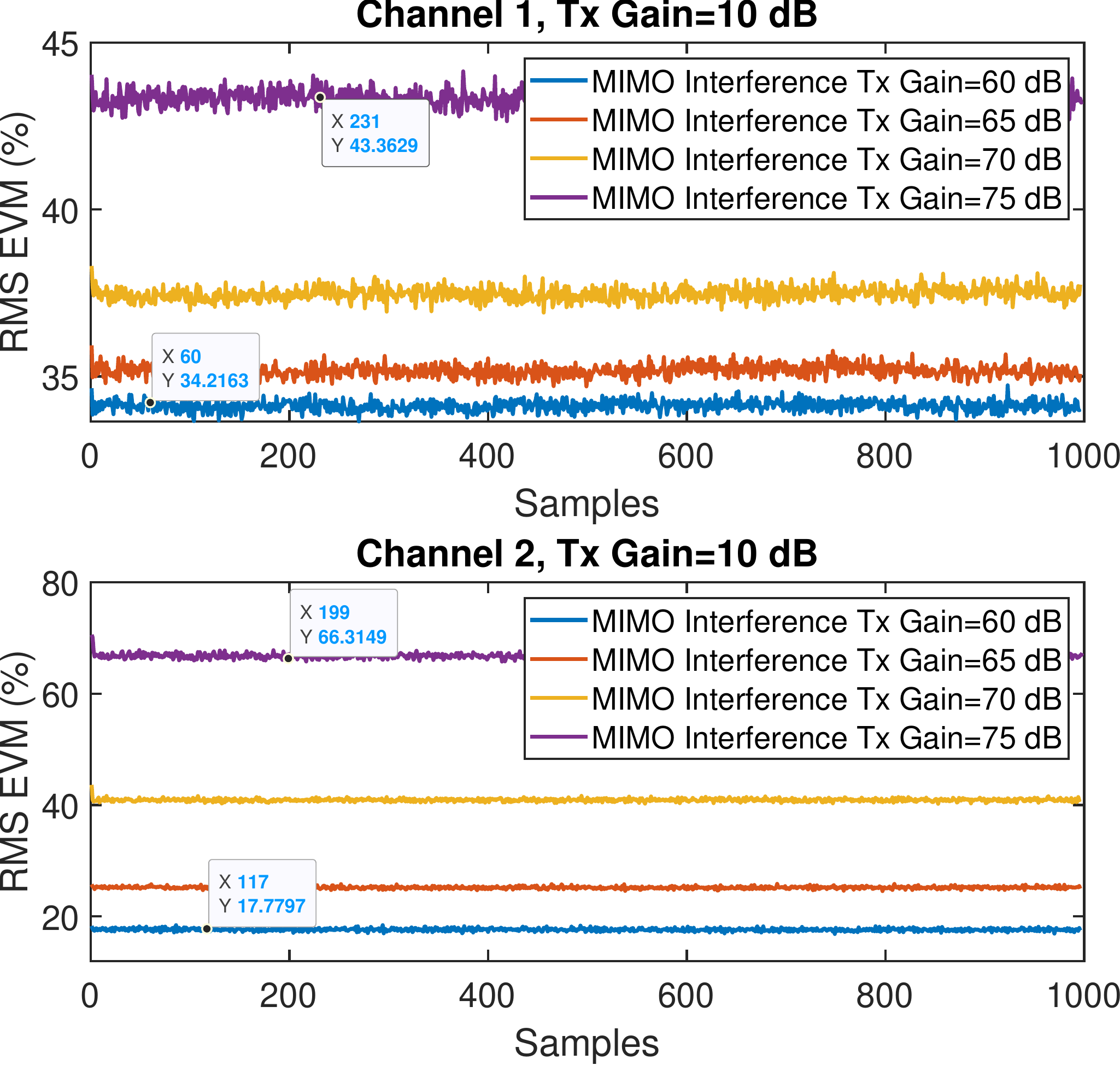}
    \caption{Measured EVM of channel 1 and channel 2 in the presence of different levels of MIMO interference. Tx gain of the USRP is 10 dB.}
    \label{fig:EVM_int_10dB}
\end{figure}

2x2 MIMO EVM measurements are presented in the presence of MIMO interference as a function of interference gain values. QPSK receiver is tested in the presence of MIMO interference generated by B210 USRP. The measurement setup is shown in Fig. \ref{fig:topview_panel_int} which shows MIMO antenna configuration and MIMO antenna interference antennas attached to the enclosure lid. B210 USRP which generated MIMO interference can also be seen on the enclosure lid. We used GNU Radio based signals and transmitted by B210 USRP as a source of interference. B210 USRP is a full-duplex fully-coherent 2x2 MIMO transceiver with integrated RF agile direct conversion transceiver AD9361 chip. B210 USRP has RF coverage from 70 MHz to 6 GHz and it supports instantaneous bandwidth of 56 MHz in 1x1 configuration and 30.72 MHz instantaneous bandwidth in 2x2 configuration. The Tx and Rx antennas are placed at a distance of 50 mm and the inter-element distance between transmit and receiving elements is 45 mm. The MIMO interference signal which is white Gaussian noise is generated using B210 USRP. Fig. \ref{fig:const_qpsk_60_int} shows QPSK constellation diagram in the presence of interference. It can be seen that the USRP channel 1 and USRP channel 2 constellation diagram degrades as the interference increases. However, the QPSK constellation diagram can be improved by increasing the Tx gain of the communication link USRP with the direct implication of increasing the signal-to-interference-plus-noise power ratio (SINR). Fig. \ref{fig:EVM_int_10dB} shows measured EVM of channel 1 and channel 2. It's clear that EVM degrades as the interference signal level increases from 60 dB to 75 dB. Channel 1 EVM has degraded from $34\%$ to $43\%$ which is $26.4\%$ increase in the RMS EVM. Similarly, channel 2 RMS EVM has degraded from $17\%$ to $66\%$ which is $288.23\%$ increase in the RMS EVM. Additionally, channel 1 RMS EVM has higher standard deviation than the channel 2 RMS EVM standard deviation.

\begin{figure}
    \centering
    \includegraphics[width=0.9\columnwidth]{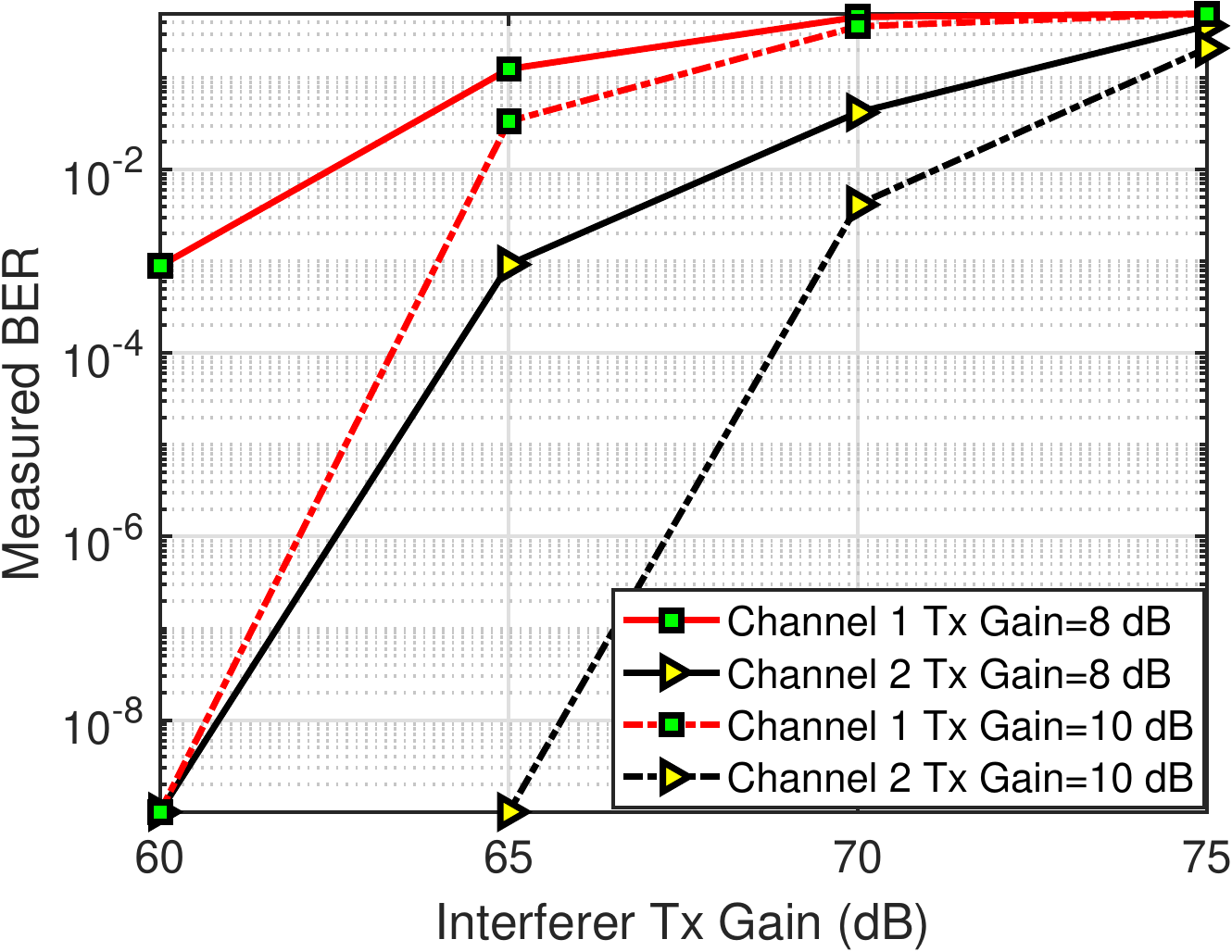}
    \caption{Measured BER of USRP channel 1 and USRP channel 2 in the presence of different MIMO interference levels.}
    \label{fig:my_label}
\end{figure}
\vspace{-0.35cm}
\subsection{Near-field MIMO BER and EVM measurements in empty and loaded enclosure}
\begin{figure*}
    \centering
    \subfloat[]{\includegraphics[width=0.72\columnwidth]{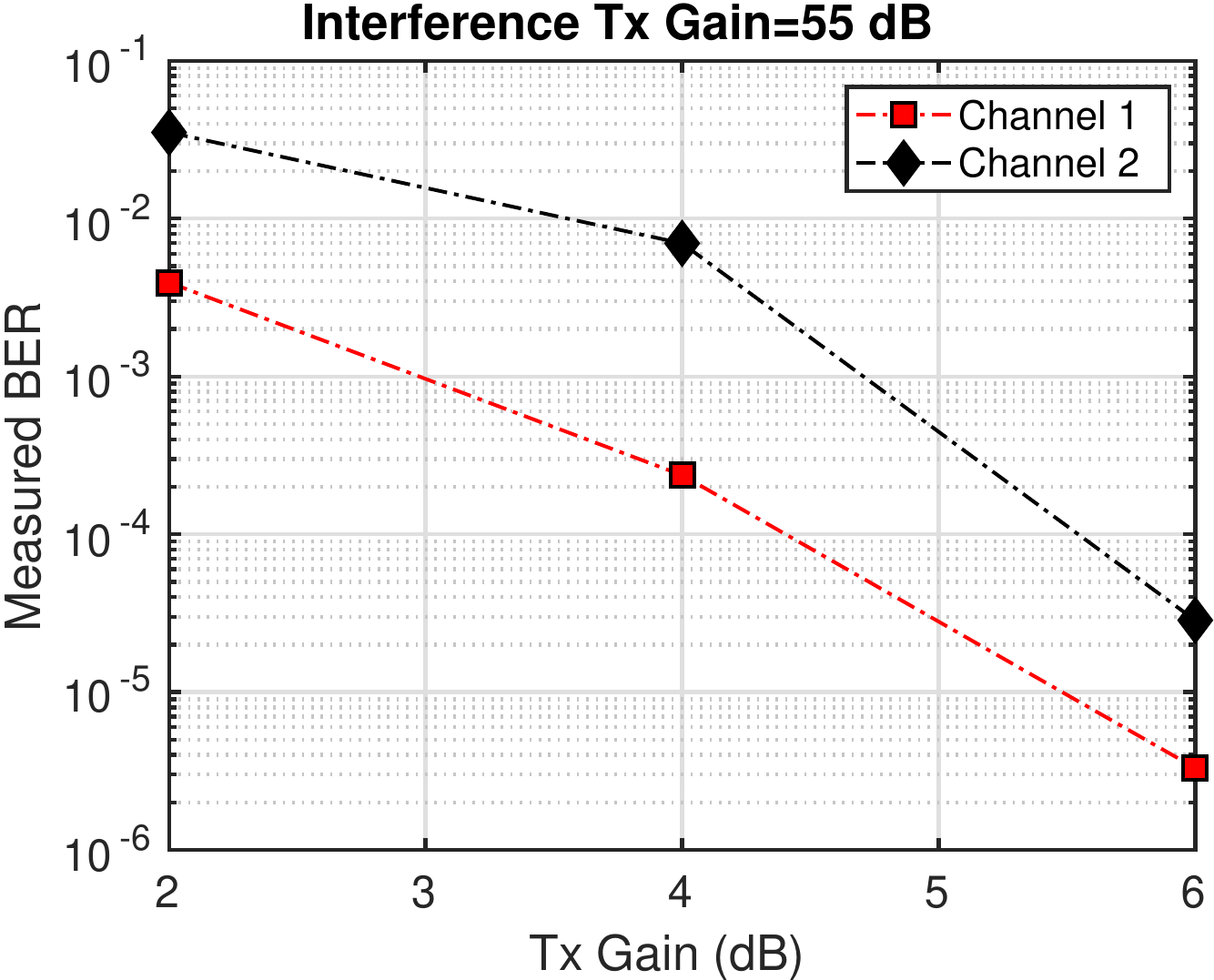}}\hspace*{2mm}
    \subfloat[]{\includegraphics[width=0.72\columnwidth]{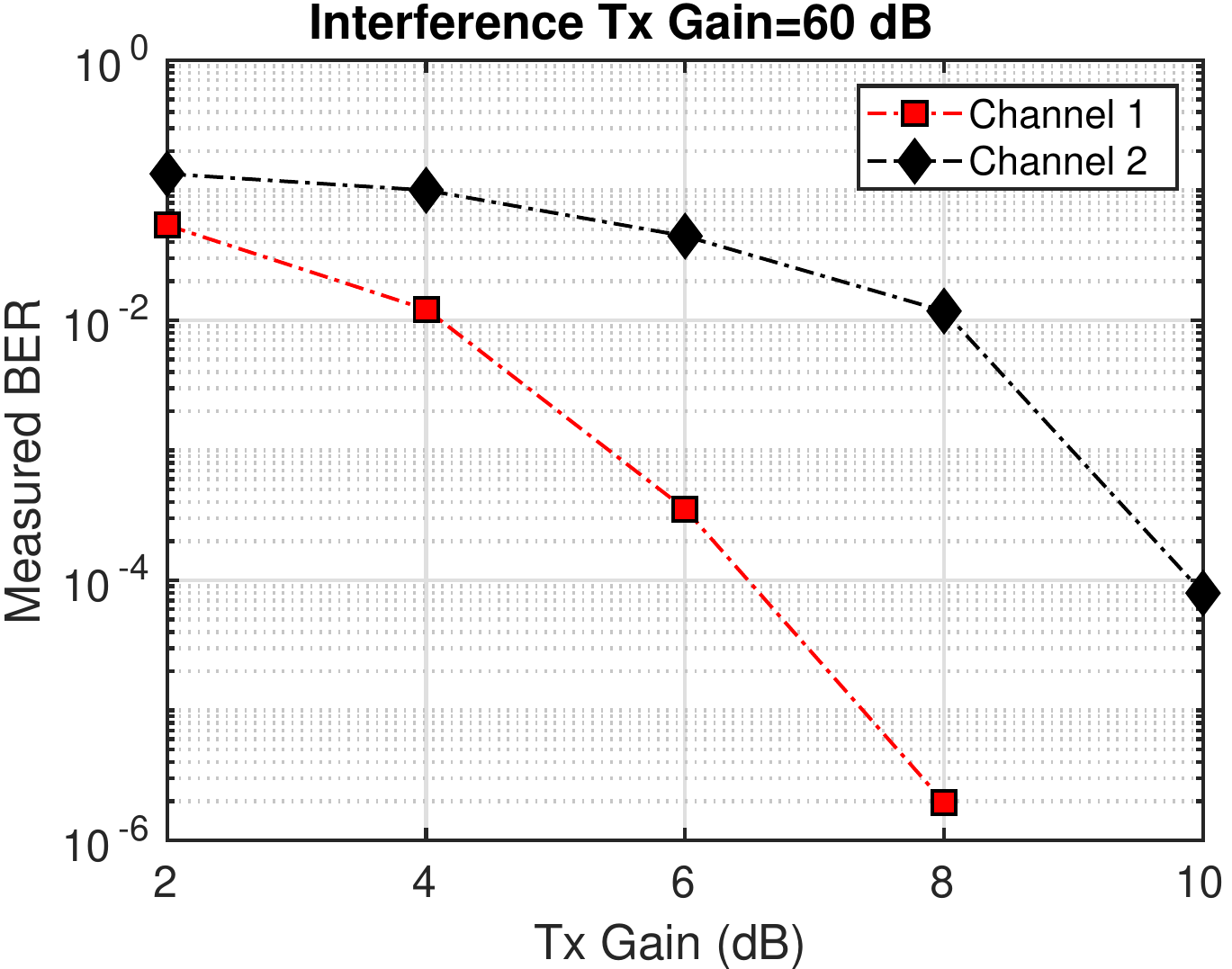}}\\
    \subfloat[interference Gain=55 dB\label{fig:evm_int55}]{\includegraphics[width=0.72\columnwidth]{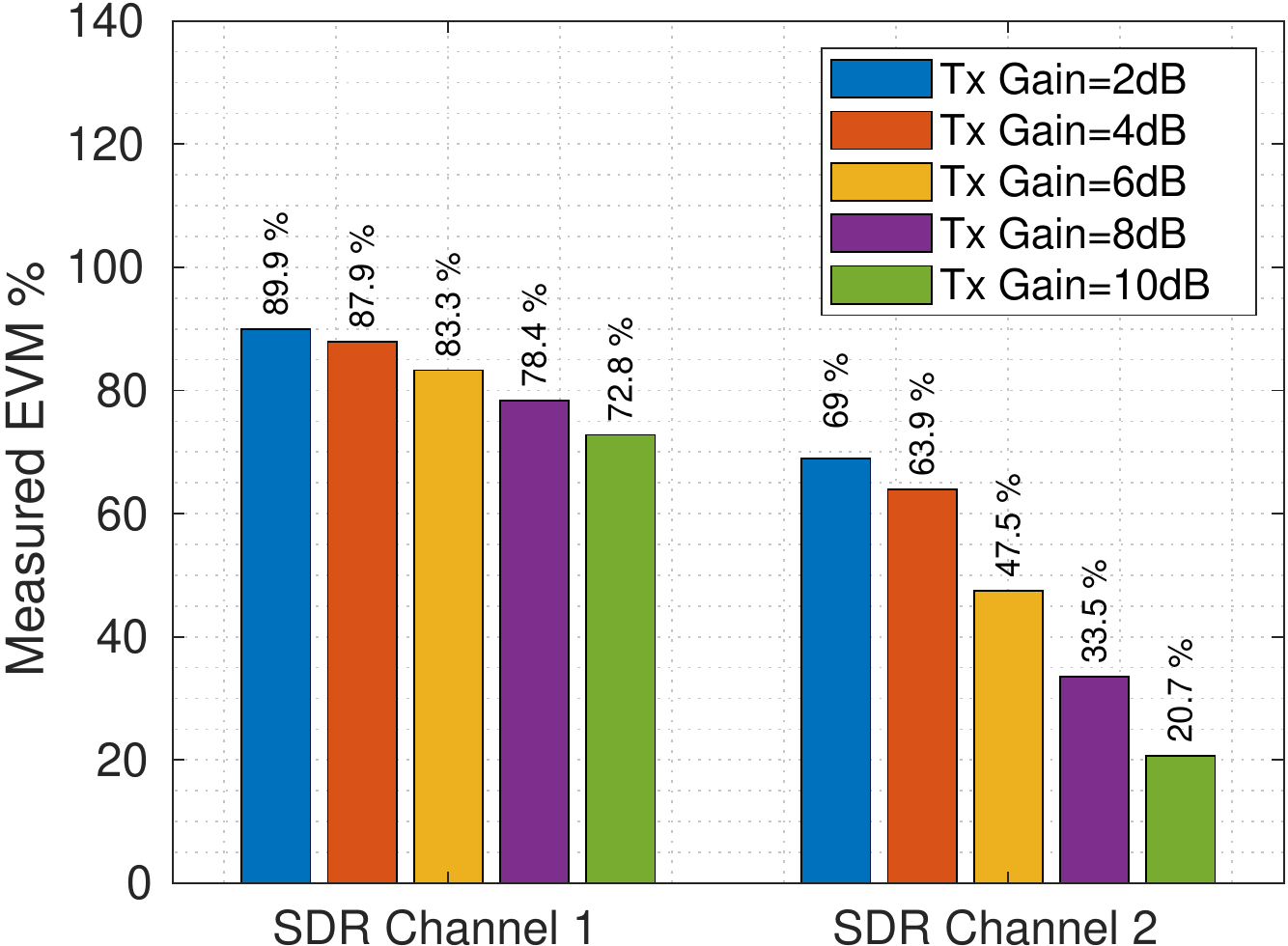}}\hspace*{2mm}
    \subfloat[interference Gain=60 dB\label{fig:evm_int60}]{\includegraphics[width=0.72\columnwidth]{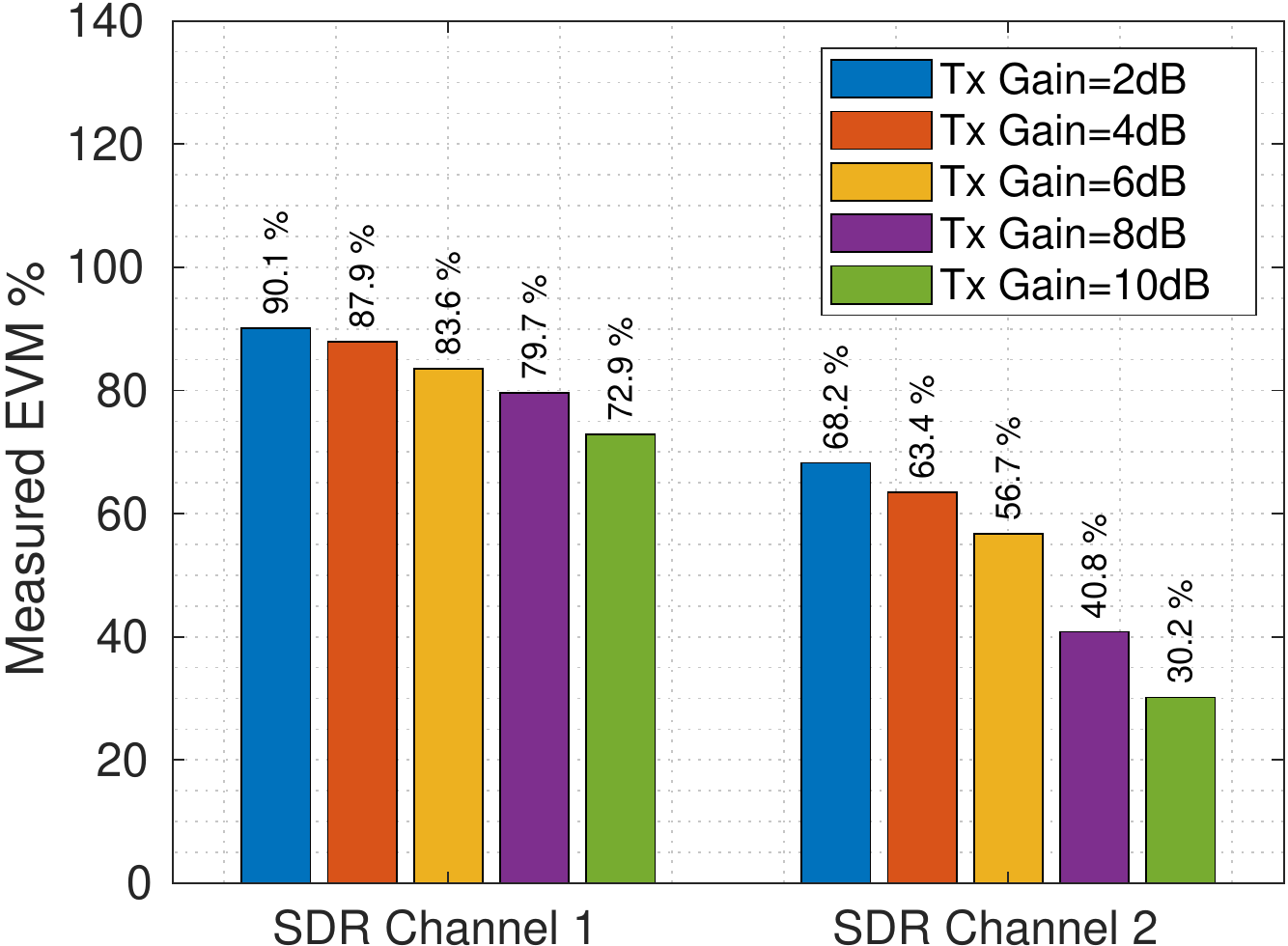}}
    \caption{Measured MIMO BER of channel 1 and channel 2 in the presence of MIMO interference when the metal enclosure was empty.}
    \label{fig:empty_enclosure_int}
\end{figure*}

In the previous section, BER and EVM were measured in the presence of interference in an empty metal enclosure. Here, the BER performance is checked in an empty enclosure and when the enclosure is loaded with losses in the form RF of absorber cones. Here BER performance of near-field MIMO communication link is measured as a function of transmit gain, enclosure losses, and interference level. Fig. \ref{fig:empty_enclosure_int} shows measured BER of channel 1 and channel 2 in an empty metal enclosure at different interference gain values. It can be seen that there are BER values at lower Tx gain values as the interference is dominant at these values for all MIMO interference values. For the all the inteferer Tx gain values the BER of two channels remain consistent i,e, channel 2 has higher BER than channel 1. BER of both the channels starts to degrade as the interference gain level increases. The BER curve for channel 1 flattens for Tx gain of 70 dB.

\begin{figure*}
    \centering
    \subfloat[]{\includegraphics[width=0.8\columnwidth]{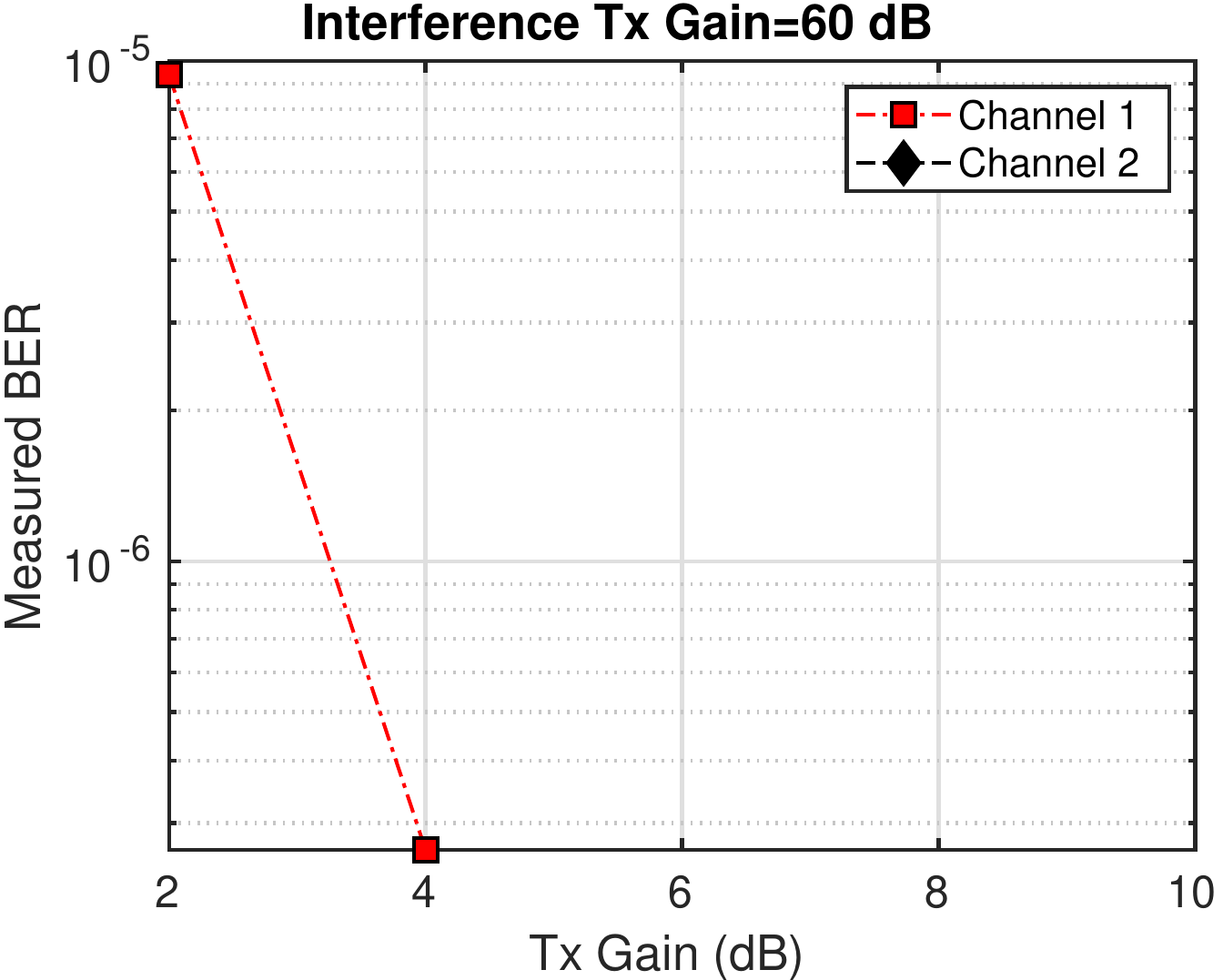}}\hspace*{2mm}
    \subfloat[]{\includegraphics[width=0.8\columnwidth]{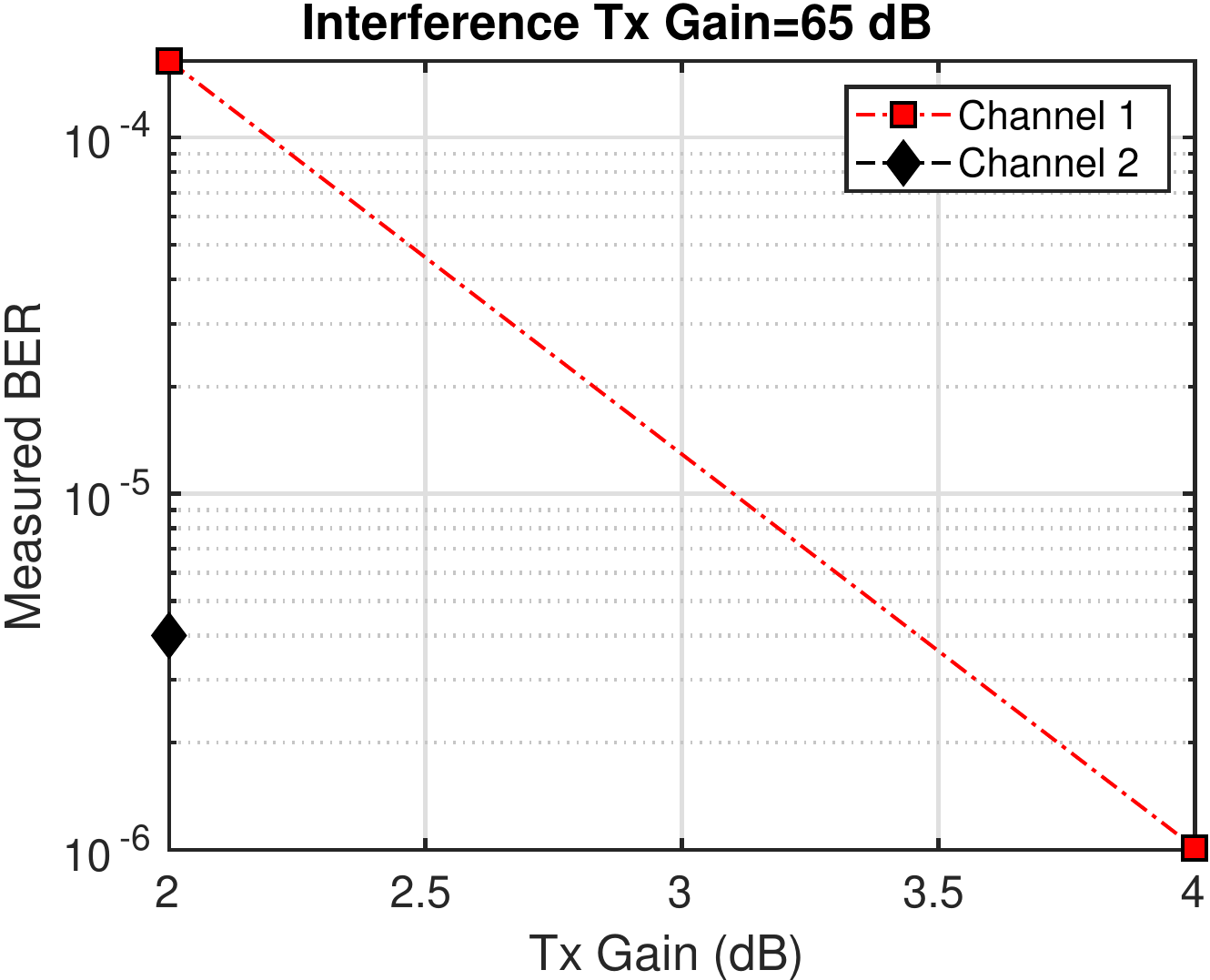}}\\
    \subfloat[]{\includegraphics[width=0.8\columnwidth]{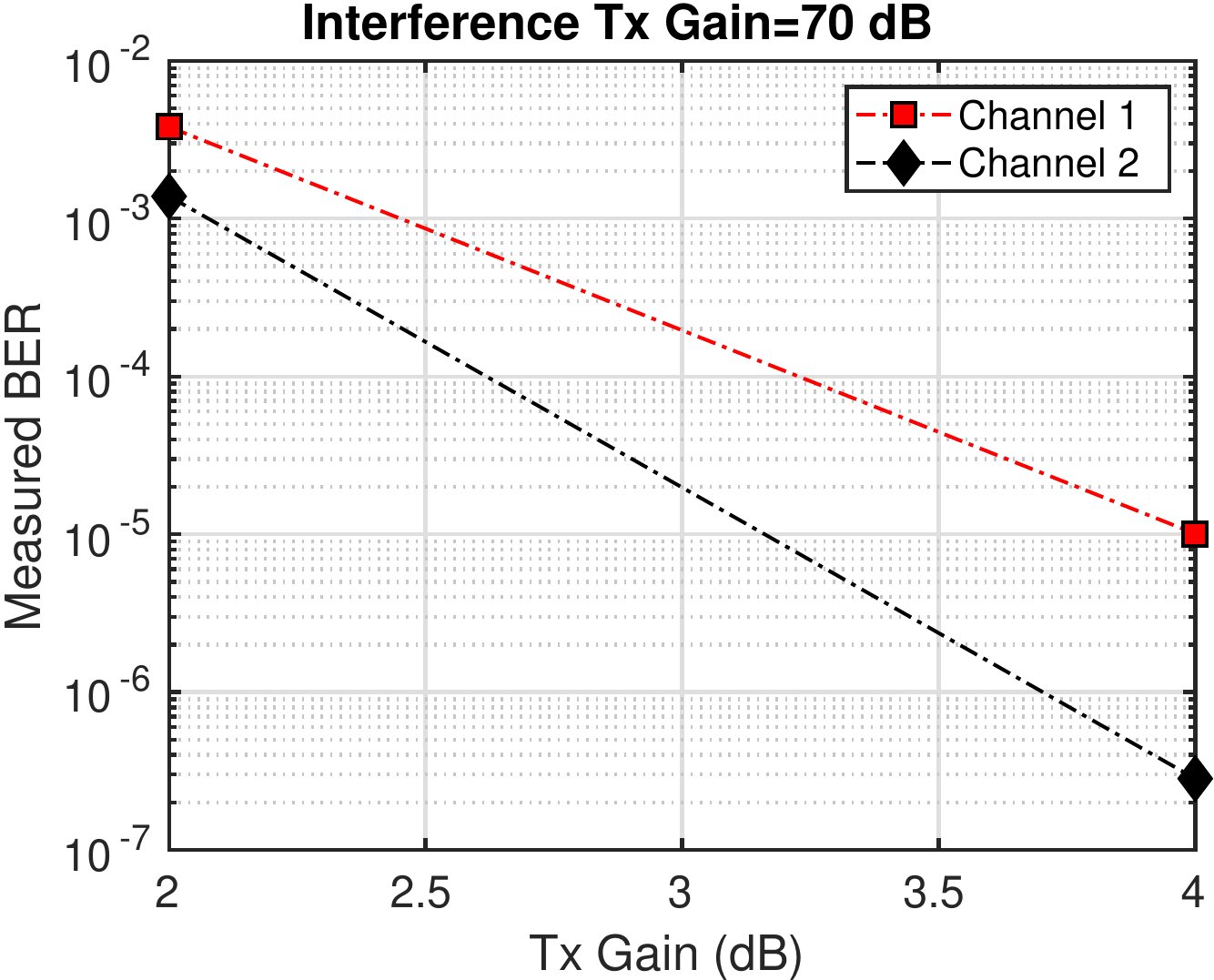}}\hspace*{2mm}
    \subfloat[]{\includegraphics[width=0.8\columnwidth]{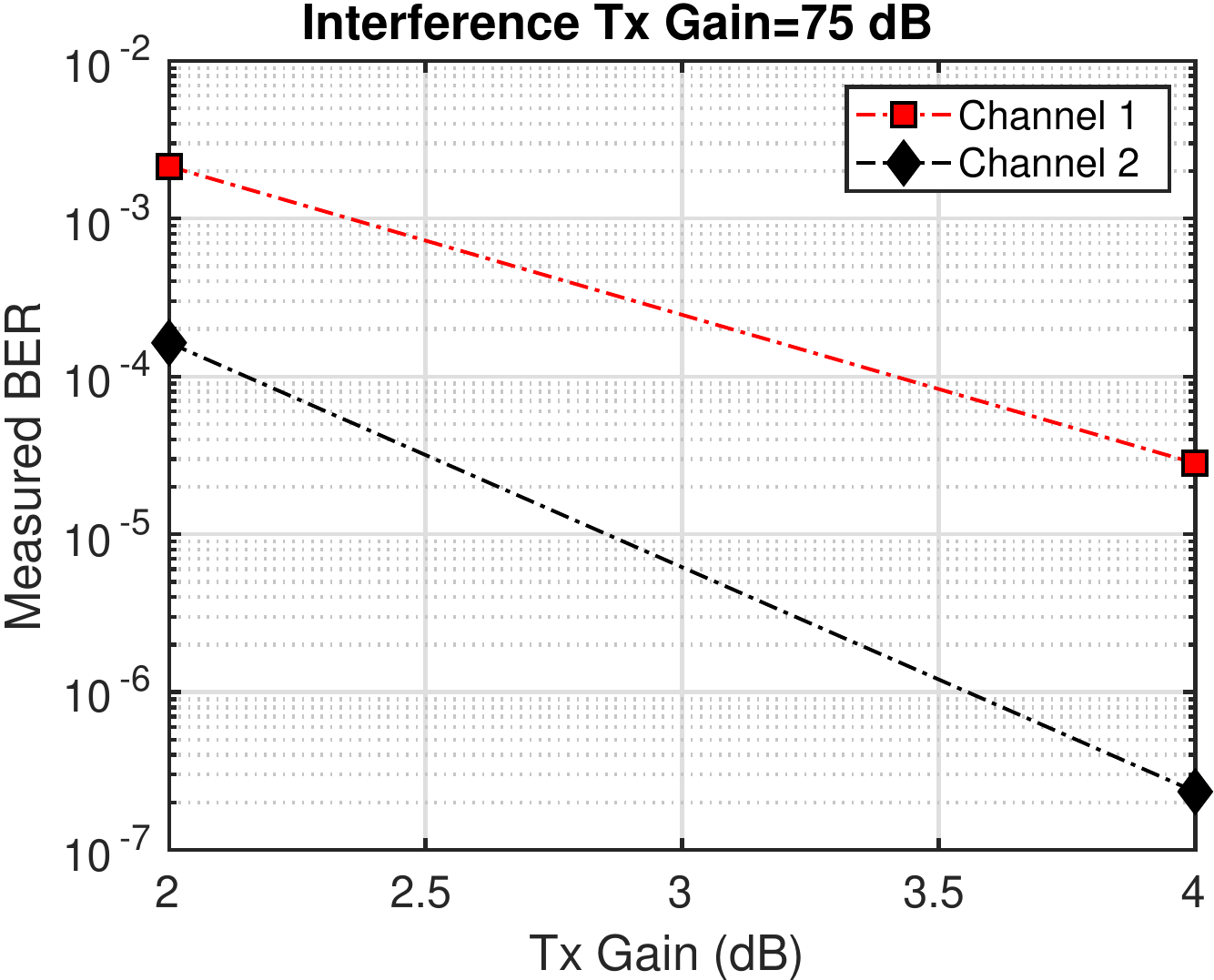}}\\
    \subfloat[]{\includegraphics[width=0.8\columnwidth]{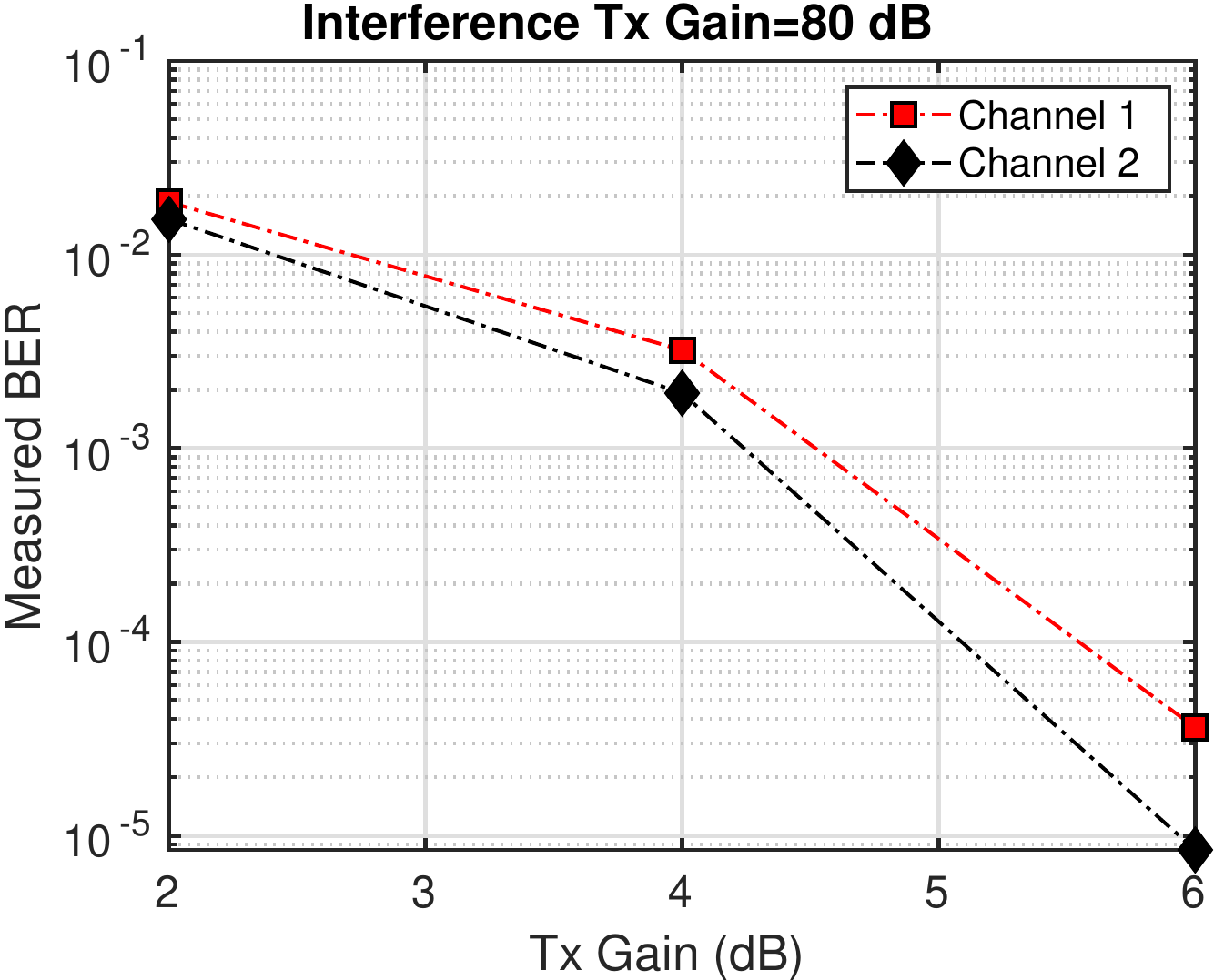}}\hspace*{2mm}
    \subfloat[]{\includegraphics[width=0.8\columnwidth]{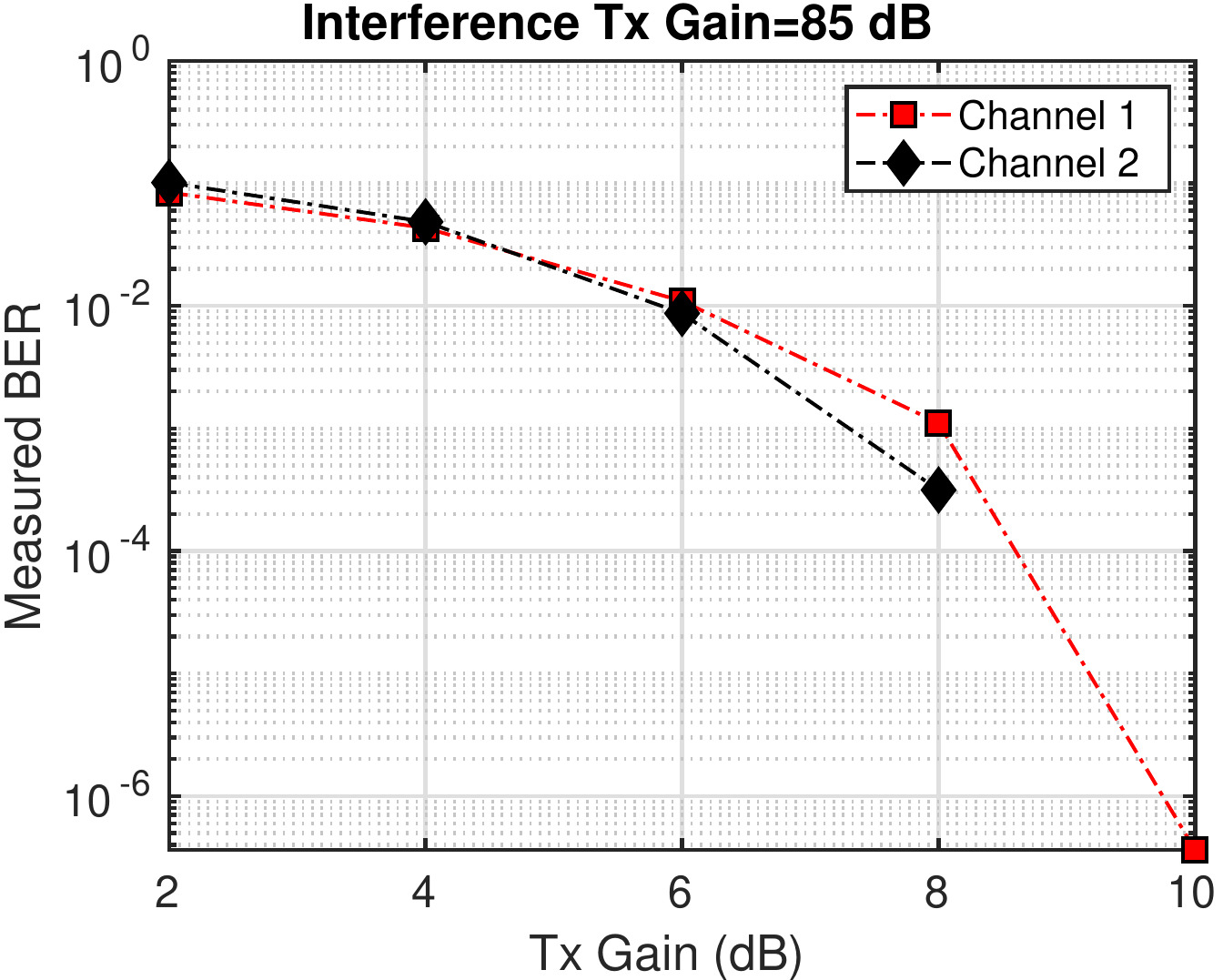}}
    \caption{Measured MIMO BER of channel 1 and channel 2 in the presence of MIMO interference when the metal enclosure was loaded with ten pieces of RF absorber cones.}
    \label{fig:enclosure_mimo_int_10cones}
\end{figure*}

\begin{figure*}
    \centering
    \subfloat[No Interference]{\includegraphics[width=0.9\columnwidth]{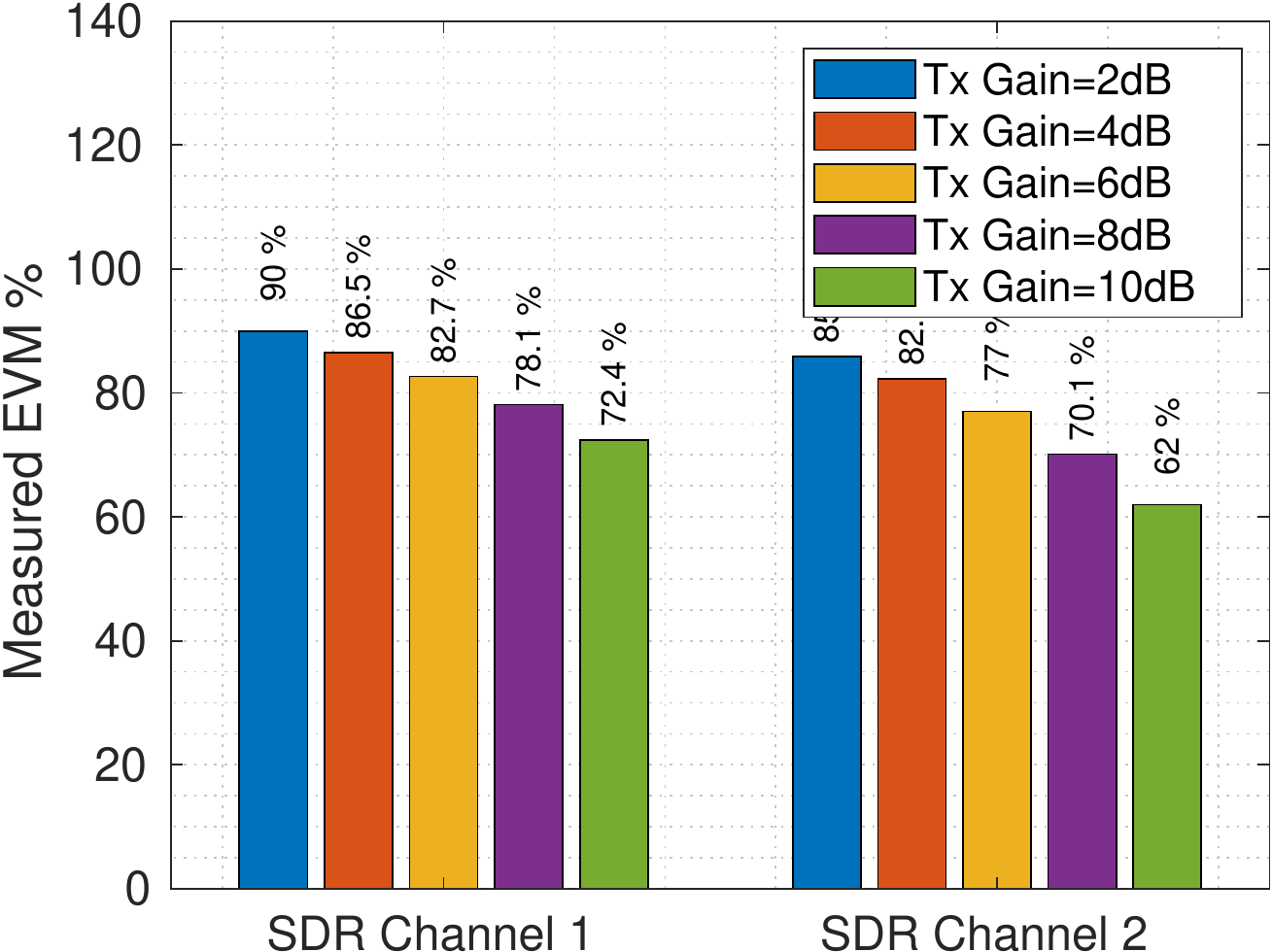}}\hspace*{2mm}
    \subfloat[Interference Gain=55 dB]{\includegraphics[width=0.9\columnwidth]{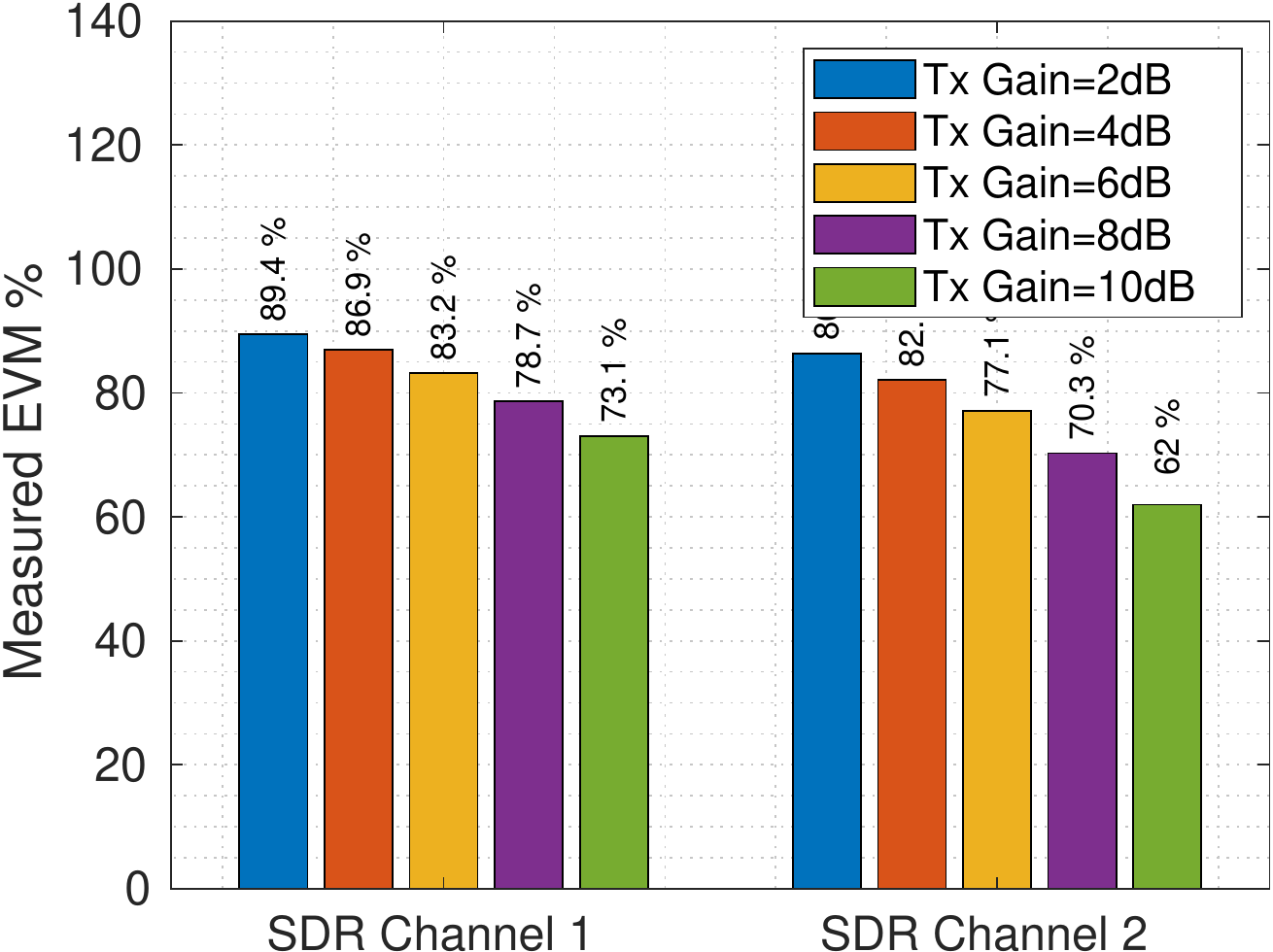}}\\
    \subfloat[Interference Gain=60 dB]{\includegraphics[width=0.9\columnwidth]{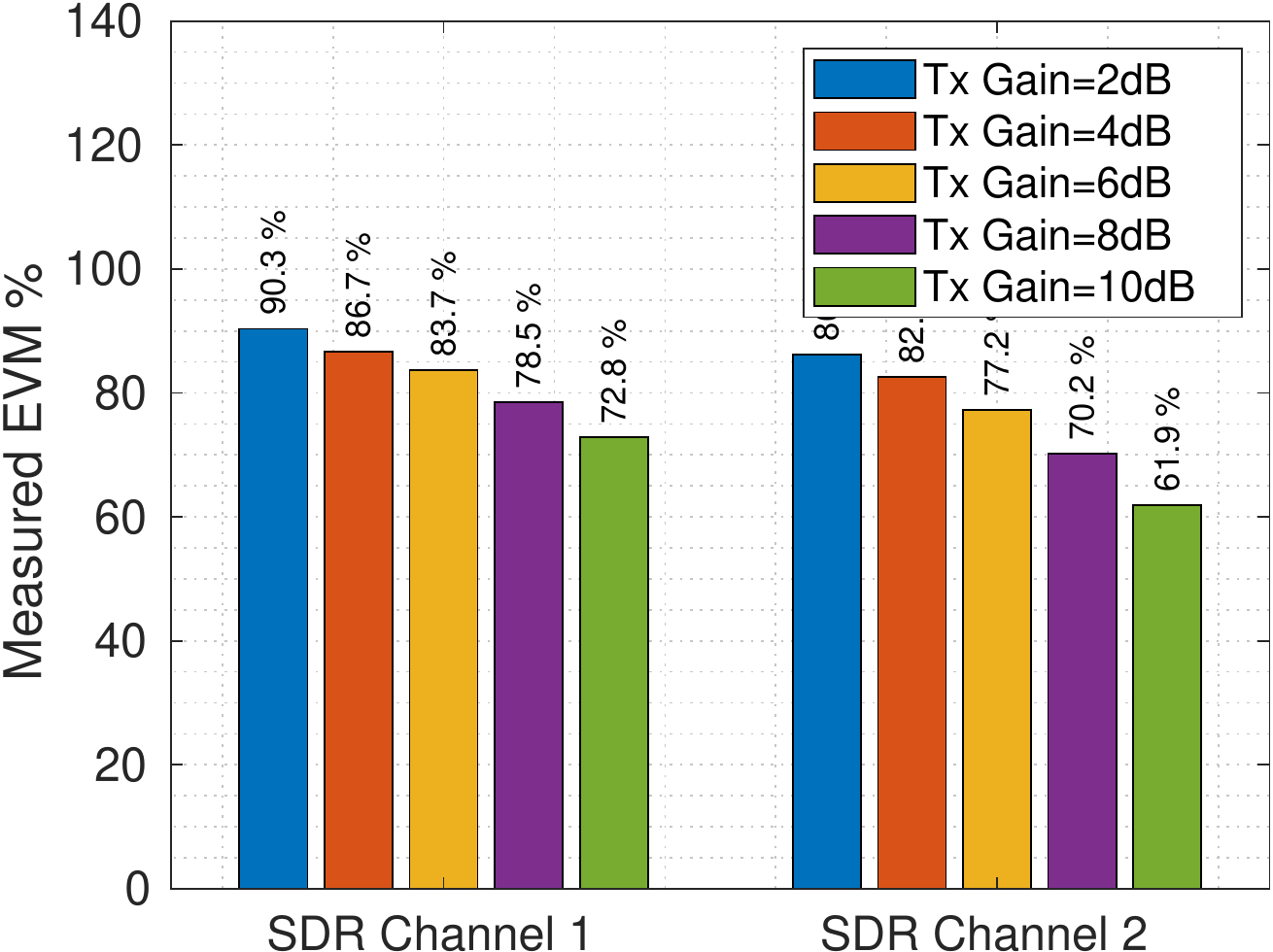}}\hspace*{2mm}
    \subfloat[Interference Gain=65 dB]{\includegraphics[width=0.9\columnwidth]{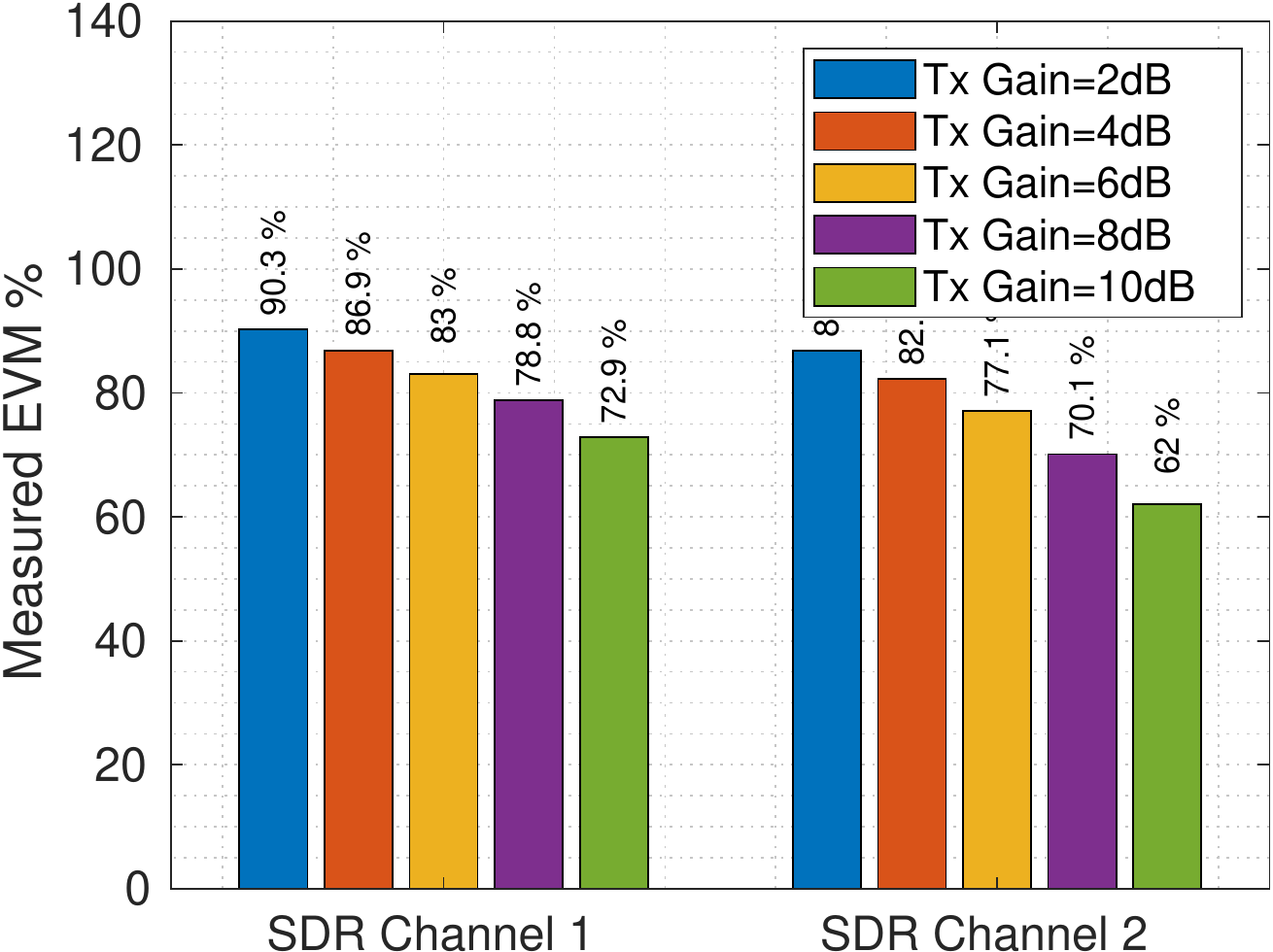}}\\
    \subfloat[Interference Gain=70 dB]{\includegraphics[width=0.9\columnwidth]{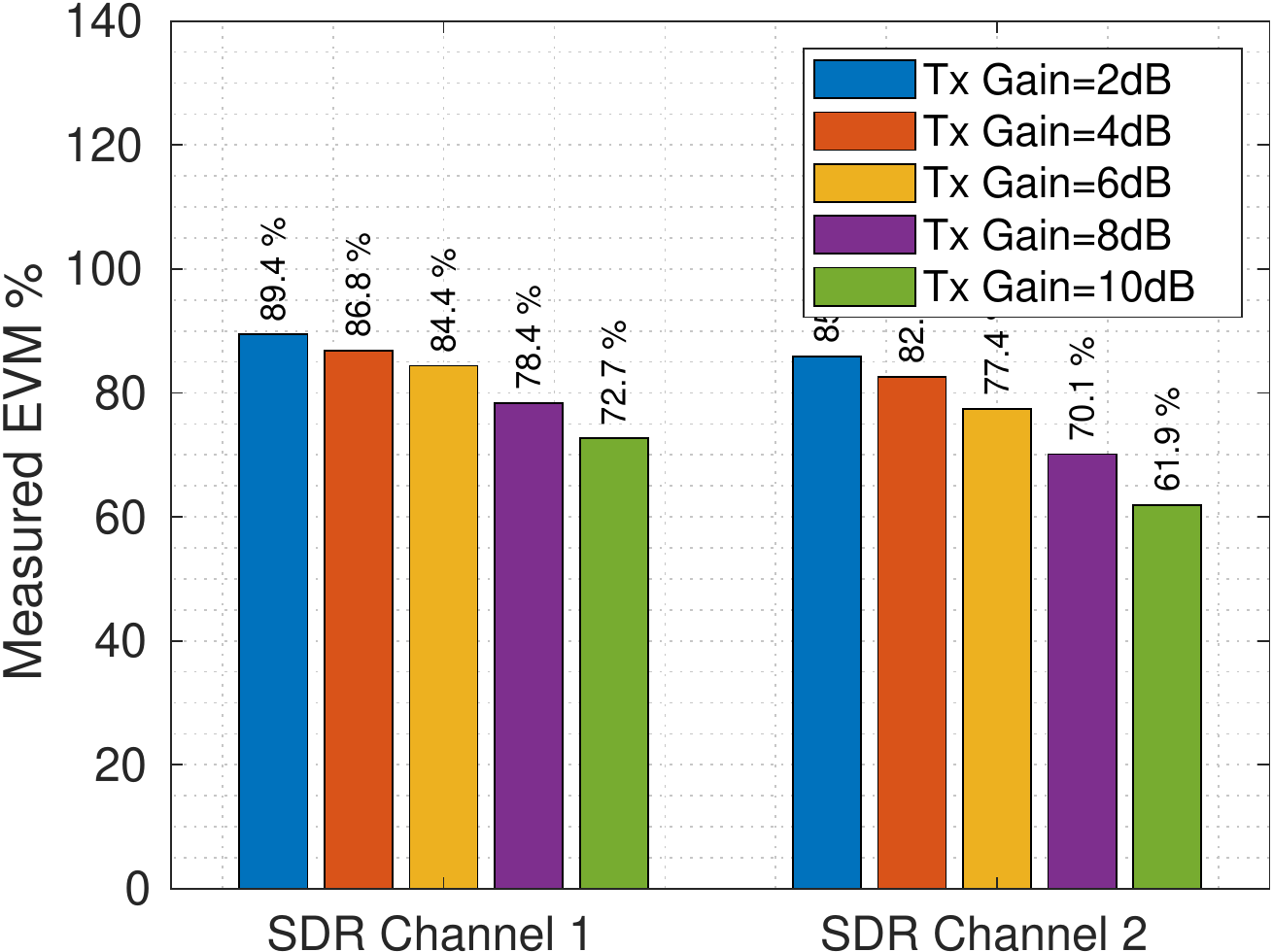}}\hspace*{2mm}
    \subfloat[Interference Gain=75 dB]{\includegraphics[width=0.9\columnwidth]{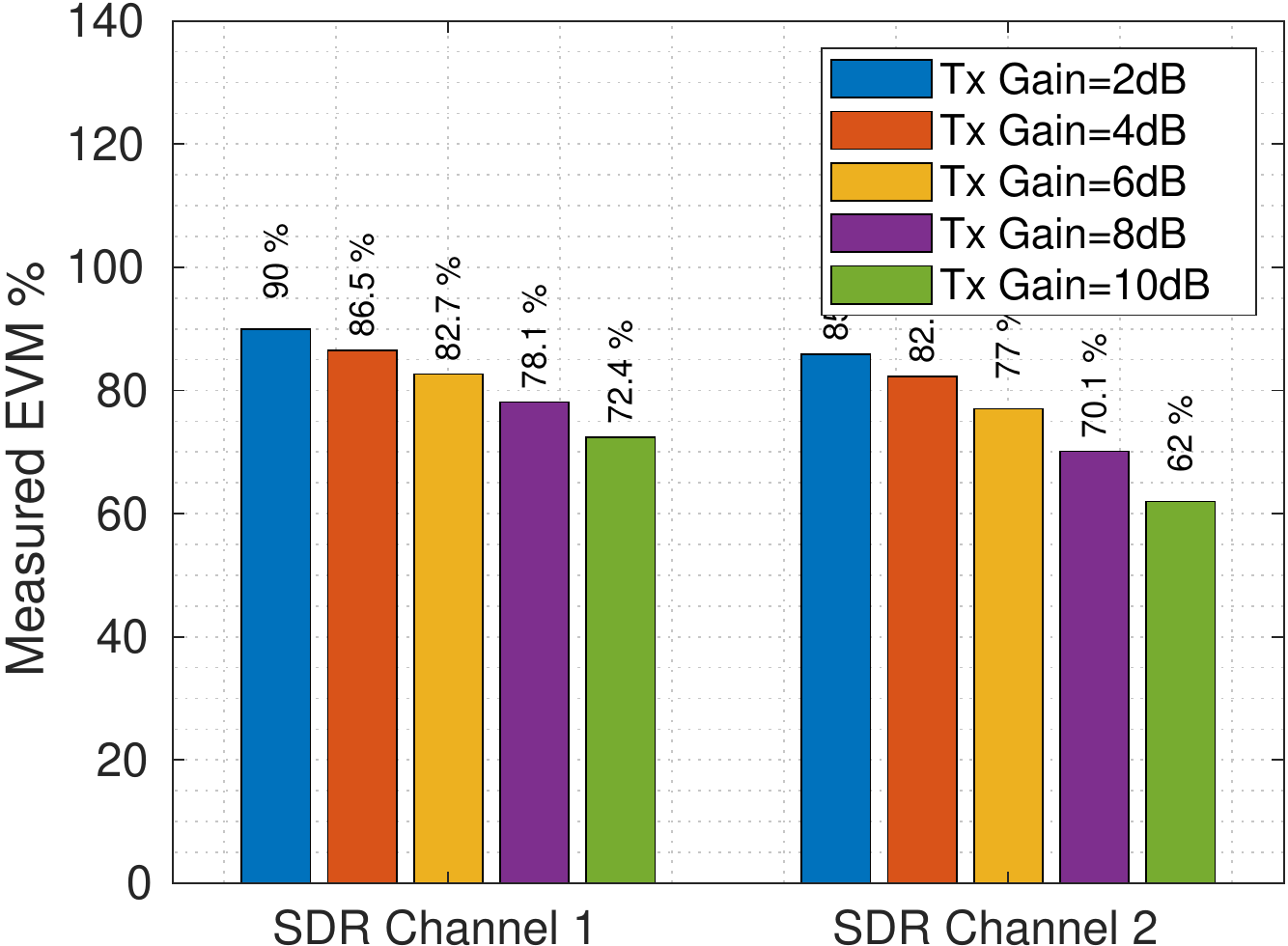}}
    \caption{Measured MIMO EVM of SDR channel 1 and SDR channel 2 in the presence of MIMO interference when the metal enclosure was loaded with ten pieces of RF absorber cones.The Tx gain was swept from 2 dB to 10 dB. AGC gain was set to 50 dB.}
    \label{fig:enclosure_mimo_50_int_EVM}
\end{figure*}

Fig.\ref{fig:evm_int55} and Fig. \ref{fig:evm_int60} shows EVM measurement of SDR channel 1 and SDR channel 2. It can be seen that EVM of channel 1 and channel 2 for all the Tx gain values at interference gain of 55 dB is less than the EVM of channel 1 and channel 2 at interference gain of 60  dB. Channel 2 EVM has degraded at high interference B210 USRP gain of 60 dB. Channel 1 EVM is degraded less than the channel 2 at all the Tx gain values. The channel 1 EVM at Tx gain of 10 dB degraded from $20.7\%$ to $30.2\%$ which is increase in the EVM by $45.89\%$. Fig. \ref{fig:enclosure_mimo_int_10cones} shows measured BER of channel 1 and channel 2 when the metal enclosure was loaded with ten RF absorber cones. It can be seen that the measured BER degrades with the increase in MIMO interference power. However, it requires higher MIMO interference power to get the same BER value of channel 1 and channel 2. Fig. \ref{fig:enclosure_mimo_50_int_EVM} shows EVM measurement of SDR channel 1 and SDR channel 2, when the empty chamber was loaded with ten pieces of RF absorbers and the AGC gain was set to 50. It can be seen that both SDR channels have comparatively stable EVM. This is because of the absorption effect of the absorbers and stabilized signal at AGC output. In the next experiment, the effect of change in AGC gain is demonstrated on EVM measurements of two USRP channels. EVM measurements are recorded by keeping the same environment and USRP configuration parameters and only changing the AGC gain from 50 dB to 60 dB.

\vspace{-0.15cm}
\section{Conclusion and Future Work}

In this work, we have shown a 2x2 near-field MIMO measurement setup and performed two-channel BER and EVM measurements in a metal enclosure. Near-field MIMO BER and EVM measurements are performed in MIMO mode using USRP in different configurations. We checked the two-channel BER and EVM measurement in a stationary environment in a metal enclosure and we found out the BER and EVM measured as a function of USRP Tx gain. In this configuration we notice BER and EVM improvement as the USRP Tx gain increased. In the second configuration, we performed BER and EVM measurements in dynamic channel conditions when the mode-stirrer moved continuously. In this scenario, we notice that the BER and EVM measurements are a function of frequency offset and dynamic channel conditions. We present the mean EVM and standard deviation of near-field EVM measurements. Two-channel EVM and BER changed dynamically as the stirrer performed movement. In the third scenario, we measured near-field MIMO BER and EVM measurements in the presence of MIMO interference. In this scenario, we observed that the BER and EVM degraded as the Tx gain of interference generated by B210 USRP increased.

\bibliographystyle{IEEEtran}

\bibliography{nf_BER.bib} 

\begin{thebibliography}{1}
\providecommand{\url}[1]{#1}
\csname url@samestyle\endcsname
\providecommand{\newblock}{\relax}
\providecommand{\bibinfo}[2]{#2}
\providecommand{\BIBentrySTDinterwordspacing}{\spaceskip=0pt\relax}
\providecommand{\BIBentryALTinterwordstretchfactor}{4}
\providecommand{\BIBentryALTinterwordspacing}{\spaceskip=\fontdimen2\font plus
\BIBentryALTinterwordstretchfactor\fontdimen3\font minus
  \fontdimen4\font\relax}
\providecommand{\BIBforeignlanguage}[2]{{%
\expandafter\ifx\csname l@#1\endcsname\relax
\typeout{** WARNING: IEEEtran.bst: No hyphenation pattern has been}%
\typeout{** loaded for the language `#1'. Using the pattern for}%
\typeout{** the default language instead.}%
\else
\language=\csname l@#1\endcsname
\fi
#2}}
\providecommand{\BIBdecl}{\relax}
\BIBdecl

\bibitem{lodro2020near}
M.~Lodro, C.~Smart, G.~Gradoni, A.~Vukovic, D.~Thomas, and S.~Greedy,
  ``Near-field ber and evm measurement at 5.8 ghz in mode-stirred metal
  enclosure.'' \emph{Applied Computational Electromagnetics Society Journal},
  vol.~35, no.~9, 2020.

\bibitem{lodro2021near}
M.~Lodro, G.~Gradoni, C.~Smartt, A.~Vukovic, D.~Thomas, and S.~Greedy,
  ``Near-field image transmission and evm measurements in rich scattering
  environment in metal enclosure,'' \emph{Progress In Electromagnetics Research
  M}, vol. 101, pp. 139--147, 2021.

\end{thebibliography}


\begin{thebibliography}{10}
\providecommand{\url}[1]{#1}
\csname url@samestyle\endcsname
\providecommand{\newblock}{\relax}
\providecommand{\bibinfo}[2]{#2}
\providecommand{\BIBentrySTDinterwordspacing}{\spaceskip=0pt\relax}
\providecommand{\BIBentryALTinterwordstretchfactor}{4}
\providecommand{\BIBentryALTinterwordspacing}{\spaceskip=\fontdimen2\font plus
\BIBentryALTinterwordstretchfactor\fontdimen3\font minus
  \fontdimen4\font\relax}
\providecommand{\BIBforeignlanguage}[2]{{%
\expandafter\ifx\csname l@#1\endcsname\relax
\typeout{** WARNING: IEEEtran.bst: No hyphenation pattern has been}%
\typeout{** loaded for the language `#1'. Using the pattern for}%
\typeout{** the default language instead.}%
\else
\language=\csname l@#1\endcsname
\fi
#2}}
\providecommand{\BIBdecl}{\relax}
\BIBdecl

\bibitem{lodro2021near}
M.~Lodro, G.~Gradoni, C.~Smartt, A.~Vukovic, D.~Thomas, and S.~Greedy,
  ``Near-field image transmission and evm measurements in rich scattering
  environment in metal enclosure,'' \emph{Progress In Electromagnetics Research
  M}, vol. 101, pp. 139--147, 2021.

\bibitem{lodro2020near}
M.~Lodro, C.~Smart, G.~Gradoni, A.~Vukovic, D.~Thomas, and S.~Greedy,
  ``Near-field ber and evm measurement at 5.8 ghz in mode-stirred metal
  enclosure.'' \emph{Applied Computational Electromagnetics Society Journal},
  vol.~35, no.~9, 2020.

\bibitem{shamim2016wireless}
M.~S. Shamim, N.~Mansoor, R.~S. Narde, V.~Kothandapani, A.~Ganguly, and
  J.~Venkataraman, ``A wireless interconnection framework for seamless inter
  and intra-chip communication in multichip systems,'' \emph{IEEE Transactions
  on Computers}, vol.~66, no.~3, pp. 389--402, 2016.

\bibitem{chen2007inter}
Z.~M. Chen and Y.~P. Zhang, ``Inter-chip wireless communication channel:
  Measurement, characterization, and modeling,'' \emph{IEEE transactions on
  antennas and propagation}, vol.~55, no.~3, pp. 978--986, 2007.

\bibitem{radi2020demonstration}
B.~Radi, A.~S. Dhillon, and O.~Liboiron-Ladouceur, ``Demonstration of
  inter-chip rf data transmission using on-chip antennas in silicon
  photonics,'' \emph{IEEE Photonics Technology Letters}, vol.~32, no.~11, pp.
  659--662, 2020.

\bibitem{timoneda2018channel}
X.~Timoneda, A.~Cabellos-Aparicio, D.~Manessis, E.~Alarc{\'o}n, and S.~Abadal,
  ``Channel characterization for chip-scale wireless communications within
  computing packages,'' in \emph{2018 Twelfth IEEE/ACM International Symposium
  on Networks-on-Chip (NOCS)}.\hskip 1em plus 0.5em minus 0.4em\relax IEEE,
  2018, pp. 1--8.

\bibitem{phang2018near}
S.~Phang, M.~T. Ivrla{\v{c}}, G.~Gradoni, S.~C. Creagh, G.~Tanner, and J.~A.
  Nossek, ``Near-field mimo communication links,'' \emph{IEEE Transactions on
  Circuits and Systems I: Regular Papers}, vol.~65, no.~9, pp. 3027--3036,
  2018.

\bibitem{kim2017review}
H.-J. Kim, H.~Hirayama, S.~Kim, K.~J. Han, R.~Zhang, and J.-W. Choi, ``Review
  of near-field wireless power and communication for biomedical applications,''
  \emph{IEEE Access}, vol.~5, pp. 21\,264--21\,285, 2017.

\bibitem{kim2016near}
H.-J. Kim, J.~Park, K.-S. Oh, J.~P. Choi, J.~E. Jang, and J.-W. Choi,
  ``Near-field magnetic induction mimo communication using heterogeneous
  multipole loop antenna array for higher data rate transmission,'' \emph{IEEE
  Transactions on Antennas and Propagation}, vol.~64, no.~5, pp. 1952--1962,
  2016.

\bibitem{guo2020performance}
H.~Guo, ``Performance analysis of near-field magnetic induction communication
  in extreme environments,'' \emph{Progress In Electromagnetics Research
  Letters}, vol.~90, pp. 77--83, 2020.

\bibitem{tan2015wireless}
X.~Tan, Z.~Sun, and I.~F. Akyildiz, ``Wireless underground sensor networks:
  Mi-based communication systems for underground applications.'' \emph{IEEE
  Antennas and Propagation Magazine}, vol.~57, no.~4, pp. 74--87, 2015.

\bibitem{wen2021channel}
E.~Wen, D.~Sievenpiper, and P.~P. Mercier, ``Channel characterization of
  magnetic human body communication,'' \emph{IEEE Transactions on Biomedical
  Engineering}, 2021.

\bibitem{akyildiz2015realizing}
I.~F. Akyildiz, P.~Wang, and Z.~Sun, ``Realizing underwater communication
  through magnetic induction,'' \emph{IEEE Communications Magazine}, vol.~53,
  no.~11, pp. 42--48, 2015.

\bibitem{li2019survey}
Y.~Li, S.~Wang, C.~Jin, Y.~Zhang, and T.~Jiang, ``A survey of underwater
  magnetic induction communications: Fundamental issues, recent advances, and
  challenges,'' \emph{IEEE Communications Surveys \& Tutorials}, vol.~21,
  no.~3, pp. 2466--2487, 2019.

\bibitem{mikki2020theory}
S.~Mikki, ``Theory of nonsinusoidal antennas for near-field communication
  system design,'' \emph{Progress In Electromagnetics Research}, vol.~86, pp.
  177--193, 2020.

\bibitem{sharma2019wideband}
A.~Sharma, E.~Kampianakis, J.~Rosenthal, A.~Pike, A.~Dadkhah, and M.~S.
  Reynolds, ``Wideband uhf dqpsk backscatter communication in reverberant
  cavity animal cage environments,'' \emph{IEEE Transactions on Antennas and
  Propagation}, vol.~67, no.~8, pp. 5002--5011, 2019.

\bibitem{ohira2011experimental}
M.~Ohira, T.~Umaba, S.~Kitazawa, H.~Ban, and M.~Ueba, ``Experimental
  characterization of microwave radio propagation in ict equipment for wireless
  harness communications,'' \emph{IEEE transactions on antennas and
  propagation}, vol.~59, no.~12, pp. 4757--4765, 2011.

\bibitem{he2017stochastic}
D.~He, K.~Guan, A.~Fricke, B.~Ai, R.~He, Z.~Zhong, A.~Kasamatsu, I.~Hosako, and
  T.~K{\"u}rner, ``Stochastic channel modeling for kiosk applications in the
  terahertz band,'' \emph{IEEE Transactions on Terahertz Science and
  Technology}, vol.~7, no.~5, pp. 502--513, 2017.

\bibitem{redfield2011understanding}
S.~Redfield, S.~Woracheewan, H.~Liu, P.~Chiang, J.~Nejedlo, and R.~Khanna,
  ``Understanding the ultrawideband channel characteristics within a computer
  chassis,'' \emph{IEEE Antennas and Wireless Propagation Letters}, vol.~10,
  pp. 191--194, 2011.

\bibitem{khademi2015channel}
S.~Khademi, S.~P. Chepuri, Z.~Irahhauten, G.~J. Janssen, and A.-J. van~der
  Veen, ``Channel measurements and modeling for a 60 ghz wireless link within a
  metal cabinet,'' \emph{IEEE Transactions on Wireless Communications},
  vol.~14, no.~9, pp. 5098--5110, 2015.

\bibitem{karedal2007characterization}
J.~Karedal, A.~P. Singh, F.~Tufvesson, and A.~F. Molisch, ``Characterization of
  a computer board-to-board ultra-wideband channel,'' \emph{IEEE Communications
  letters}, vol.~11, no.~6, pp. 468--470, 2007.

\bibitem{gelabert2011experimental}
J.~Gelabert, D.~Edwards, and C.~Stevens, ``Experimental evaluation of uwb
  wireless communication within pc case,'' \emph{Electronics Letters}, vol.~47,
  no.~13, pp. 773--775, 2011.

\end{thebibliography}
\end{document}